\documentclass[conference]{IEEEtran}
\usepackage{cite}
\usepackage{amsmath,amssymb,amsfonts}
\usepackage{algorithmic}
\usepackage{graphicx}
\usepackage{textcomp}
\usepackage{xcolor}
\usepackage[hyphens]{url}
\usepackage{fancyhdr}
\usepackage{hyperref}
\usepackage{pifont}
\usepackage{subcaption}
\usepackage{multirow}
\usepackage{makecell}
\usepackage{arydshln}

\newcommand{\squishlist}{
 \begin{list}{$\bullet$}
  { \setlength{\itemsep}{0pt}
     \setlength{\parsep}{0pt}
     \setlength{\topsep}{3pt}
     \setlength{\partopsep}{0pt}
     \setlength{\leftmargin}{1.5em}
     \setlength{\labelwidth}{1em}
     \setlength{\labelsep}{0.5em} } }

\newcommand{\squishnums}{
 \begin{list}{$\bullets$}
  { \setlength{\itemsep}{0pt}
     \setlength{\parsep}{3pt}
     \setlength{\topsep}{3pt}
     \setlength{\partopsep}{0pt}
     \setlength{\leftmargin}{1.5em}
     \setlength{\labelwidth}{1em}
     \setlength{\labelsep}{0.5em} } }

\newcommand{\squishlisttwo}{
 \begin{list}{$\bullet$}
  { \setlength{\itemsep}{0pt}
     \setlength{\parsep}{0pt}
    \setlength{\topsep}{0pt}
    \setlength{\partopsep}{0pt}
    \setlength{\leftmargin}{2em}
    \setlength{\labelwidth}{1.5em}
    \setlength{\labelsep}{0.5em} } }

\newcommand{\squishend}{
  \end{list}  }

\newcommand{\squishnobullet}{
 \begin{list}{}
  { \setlength{\itemsep}{0pt}
     \setlength{\parsep}{0pt}
     \setlength{\topsep}{3pt}
     \setlength{\partopsep}{0pt}
     \setlength{\leftmargin}{0.5em}
     \setlength{\labelwidth}{1em}
     \setlength{\labelsep}{0.5em} } }

\newcommand{\secref}[1]{\S\ref{#1}}
\newcommand{\figref}[1]{Fig.~\ref{#1}}
\newcommand{\tabref}[1]{Tab.~\ref{#1}}

\newif\ifcommenton
\commentontrue
\ifcommenton
\newcommand{\TODO}[1]{\textcolor{red}{[TODO] #1}}
\newcommand{\JT}[1]{{\color{brown}\bfseries [Jianming: #1]}}
\newcommand{\AI}[1]{{\color{blue}\bfseries [Anirudh: #1]}}
\newcommand{\PC}[1]{{\color{blue}\bfseries [Prasanth: #1]}}
\newcommand{\TK}[1]{{\color{violet}\bfseries [TK: #1]}}
\newcommand{\GJ}[1]{{\color{blue}\bfseries [GJ: #1]}}
\newcommand{\THH}[1]{{\color{teal} [Tianhao: #1]}}
\newcommand{\DVJ}[1]{{\color{cyan}\bfseries [DJ: #1]}}
\newcommand{\fixme}[1]{{{\color{blue} #1}}}
\else
\newcorowcolormmand{\TODO}[1]{}
\newcommand{\AI}[1]{}
\newcommand{\PC}[1]{}
\newcommand{\JT}[1]{}
\newcommand{\TK}[1]{}
\newcommand{\THH}[1]{}
\newcommand{\GJ}[1]{}
\newcommand{\DVJ}[1]{}
\newcommand{\fixme}[1]{}
\fi

\definecolor{codegreen}{rgb}{0,0.6,0}
\definecolor{codegray}{rgb}{0.5,0.5,0.5}
\definecolor{codepurple}{rgb}{0.58,0,0.82}
\definecolor{backcolour}{rgb}{0.95,0.95,0.92}
\definecolor{purple}{RGB}{128,0,128}
\definecolor{indigo}{RGB}{75,0,130}
\definecolor{royalblue}{RGB}{65,105,225}
\definecolor{navy}{RGB}{0,0,128}
\definecolor{myred}{RGB}{153,0,0}
\definecolor{myblue}{RGB}{0,76,153}
\definecolor{mygreen}{RGB}{0,102,51}
\definecolor{mypurple}{RGB}{76,0,153}

\newcommand{\ivn}{{\color{myblue}{$I_{VN}$}}}
\newcommand{\wvn}{{\color{mygreen}{$W_{VN}$}}}
\newcommand{\swvn}{{\color{mygreen}{W_{VN}}}}
\newcommand{\ovn}{{\color{myred}{$O_{VN}$}}}
\newcommand{\pvn}{{\color{mypurple}{$P_{VN}$}}}

\fancypagestyle{camerareadyfirstpage}{%
  \fancyhead{}
  
  \fancyhead[C]{
    \ifdefined\aeopen
    \parbox[][12mm][t]{13.5cm}{2026 IEEE International Symposium on Performance Analysis of Systems and Software}    
    \else
      \ifdefined\aereviewed
      \parbox[][12mm][t]{13.5cm}{2026 IEEE International Symposium on Performance Analysis of Systems and Software (ISPASS)}
      \else
      \ifdefined\aereproduced
      \parbox[][12mm][t]{13.5cm}{2026 IEEE International Symposium on Performance Analysis of Systems and Software (ISPASS)}
      \else
      \parbox[][0mm][t]{13.5cm}{2026 IEEE International Symposium on Performance Analysis of Systems and Software (ISPASS)}
    \fi 
    \fi 
    \fi 
    \ifdefined\aeopen 
      \includegraphics[width=12mm,height=12mm]{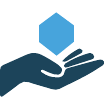}
    \fi 
    \ifdefined\aereviewed
      \includegraphics[width=12mm,height=12mm]{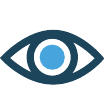}
    \fi 
    \ifdefined\aereproduced
      \includegraphics[width=12mm,height=12mm]{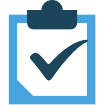}
    \fi
  }
  \fancyfoot[C]{}
}

\def\aeopen{}           % The artifact is publically available
\def\aereviewed{}     % The artefact has been reviewed
\def\aereproduced{} % The results have been reproduced

\def\BibTeX{{\rm B\kern-.05em{\sc i\kern-.025em b}\kern-.08em
    T\kern-.1667em\lower.7ex\hbox{E}\kern-.125emX}}
\begin{document}
\pagestyle{plain}
\title{MINISA: Minimal Instruction Set Architecture for Next-gen Reconfigurable Inference Accelerator \\
}

\author{
\IEEEauthorblockN{Jianming Tong\IEEEauthorrefmark{3}, Devansh Jain\IEEEauthorrefmark{2}, Yujie Li\IEEEauthorrefmark{3}, Charith Mendis\IEEEauthorrefmark{2}, Tushar Krishna\IEEEauthorrefmark{3}}
\IEEEauthorblockA{\IEEEauthorrefmark{2}\textit{University of Illinois}, Urbana-Champaign, USA \\devansh9@illinois.edu, charithm@illinois.edu}
\IEEEauthorblockA{\IEEEauthorrefmark{3}\textit{Georgia Institute of Technology}, Atlanta, Georgia, USA \\jianming.tong@gatech.edu, yli3821@gatech.edu, tushar@ece.gatech.edu}
}

\maketitle
\thispagestyle{camerareadyfirstpage}
\begin{abstract}
\label{sec:abstract}
Modern reconfigurable AI accelerators rely on rich mapping and data-layout flexibility to sustain high utilization across matrix multiplication, convolution, and emerging applications beyond AI. However, exposing this flexibility through fine-grained micro-control results in prohibitive control overhead of fetching configuration bits from off-chip memory. This paper presents MINISA, a minimal instruction set that programs a reconfigurable accelerator at the granularity of Virtual Neurons (VNs), the coarsest control granularity that retains flexibility of hardware and the finest granularity that avoids unnecessary control costs. First, we introduce FEATHER+, a modest refinement of FEATHER, that eliminates redundant on-chip replication needed for runtime dataflow/layout co-switching and supports dynamic cases where input and weight data are unavailable before execution for offline layout manipulation. MINISA then abstracts control of FEATHER+ into three layout-setting instructions for input, weight, and output VNs and a single mapping instruction for setting dataflow. This reduces the control and instruction footprint while preserving the legal mapping and layout space supported by the FEATHER+. Our results show that MINISA reduces geometric mean off-chip instruction traffic by factors ranging from $35{\times}$ to $(4\times10^5){\times}$ under various sizes under 50 GEMM workloads spanning AI (GPT-oss), FHE, and ZKP. This eliminates instruction-fetch stalls that consume $96.9\%$ of micro-instruction cycles, yielding up to $31.6\times$ end-to-end speedup for $16\times256$ FEATHER+. Our code: \url{https://github.com/maeri-project/FEATHER/tree/main/minisa}.
\end{abstract}

\section{Introduction}
\label{sec:introduction}
The new AI revolution drives crazily increasing installs of data centers for AI. With AI workloads become diverse, AI accelerators increasingly become more reconfigurable for efficiently support workloads of diverse shapes. Their performance stems from two sources: (i) high-throughput low-precision matrix/vector execution and (ii) programmable data movement that enables diverse dataflows and layouts. This programmability is now crucial not only for CNNs and Transformers, but also for workloads that can be compiled into matrix and vector operators, including homomorphic encryption (HE\cite{tong2025CROSS}) and zero-knowledge proofs (ZKP\cite{tong2025MORPH}). Yet this same flexibility creates a challenge: \textbf{control overhead}.

FEATHER\cite{tong2024FEATHER} represents the SotA accelerator enabling (dataflow, layout) coswitching for each individual workload. However, its programming interfaces expose \textbf{fine-grained} or \textbf{micro-coded} control over buffer layouts, on-chip routing, and data-to-PE mappings. While expressive, such a paradigm inflates instruction traces, increases off-chip instruction fetch traffic, and consumes limited valuable on-chip memory for storing instructions. The resulting control footprint can directly limit the size of compute tiles that fit on the chip, reducing arithmetic intensity and hurting end-to-end throughput. As accelerators scale in array dimensions and workload diversity, this control overhead becomes a first-order bottleneck (\secref{sec:overhead}).

This paper argues that the right abstraction boundary for minimizing control overhead is the \textbf{Virtual Neuron (VN)}. A VN corresponds to the smallest hardware dot-product atom. A \textbf{coarser-grained} abstraction than current VN would ignore inter-PE mapping flexibility, while a \textbf{finer-grained} abstraction would introduce unnecessary control costs without gaining flexibility. Therefore, programming at the level of VN is exactly the level to minimize control overhead without losing hardware built-in flexibility.

Building on this insight, we present a compact VN-abstracted instruction set architecture, MINISA, that captures the flexibility in mapping and layout of FEATHER\cite{tong2024FEATHER}, one SotA reconfigurable accelerator, using only eight instructions. Layout-setting instructions parameterize the nested-loop order of VNs in on-chip buffers. The mapping instruction specifies dataflow and triggers execution of a compute tile under the current layout configurations. For a single layer, the program trace consists of three layout instructions followed by a sequence of \texttt{ExecuteMapping} and \texttt{ExecuteStreaming} invocations. For consecutive layers, the output of the first layer becomes the input of the next, allowing \texttt{SetIVNLayout} of the next layer to be used to specify the output layout of next one. This allows skipping \texttt{SetOVNLayout} of the current one.

While MINISA compresses FEATHER’s control footprint, the baseline FEATHER microarchitecture still creates two practical frictions for dynamic workloads: it can induce data duplication in on-chip buffers, and its buffer/interconnect design relies on the assumption that one operand (often weights) needs to be pre-known and pre-reordered into preferred layout before program execution. This assumption limits applicability when both operands may arrive or change at runtime, such as LLM inference. To remove these barriers without redesigning the accelerator, we augment FEATHER\cite{tong2024FEATHER} into FEATHER+ with minimal architectural refinements on (1) interconnect and (2) buffer organization to add distribution flexibility for (1) eliminating data duplication in on-chip buffers, and (2) supporting scenarios when weights are not pre-known for compile-time layout pre-reorder. Our contributions are:

\noindent $\bullet$ We identify control overhead as a key limiter of tile size and instruction efficiency in reconfigurable AI accelerators, instruction fetching stall share rises from 0\% (FEATHER+ with smaller than $8\times8$ PE Array) to 96.9\% (16$\times$256 FEATHER+).

\noindent $\bullet$ We propose FEATHER+, a modest enhancement that (1) supports dynamic input data, (2) removes data duplication in on-chip buffers, and (3) enables inputs to be loaded inside NEST and streaming weights.

\noindent $\bullet$ We introduce MINISA, a minimal VN-level ISA for FEATHER that preserves hardware-supported mapping and layout flexibility with eight instructions.

\noindent $\bullet$ We demonstrate that MINISA enables up-to $31.6\times$ speedups by avoiding instruction-fetch stalls. We show MINISA reduces instruction-to-data ratio from $\sim$100$\times$ (micro-instruction programming) to negligible (MINISA) for 16$\times$256 FEATHER+. MINISA maintains $<0.1\%$ instruction-cycle fraction. %We demonstrate reduced off-chip instruction traffic and improved on-chip capacity utilization, enabling larger tiles and higher end-to-end performance across AI, HE, and ZKP kernels.

\section{Background}
\label{sec:motivation}
\begin{figure}[!t]
    \centering
    % \vspace{-3mm}
    \includegraphics[width=\columnwidth]{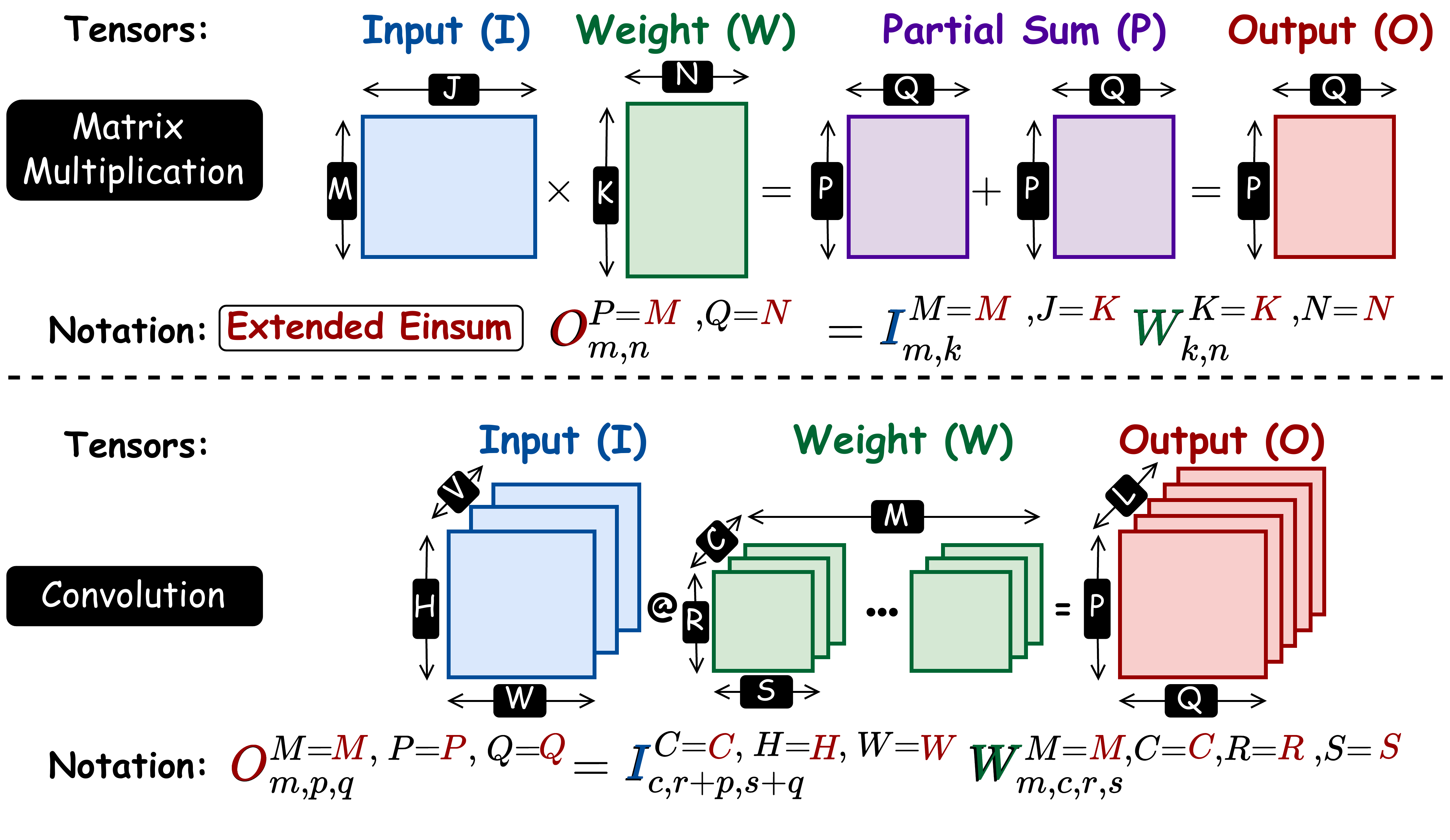}
    \vspace{-6mm}
    \caption{Workload Illustration. Convolution is converted to MatMul via im2col. We use color coding: blue/green/purple/red for Input({\color{myblue}$I$})/Weight({\color{mygreen}$W$})/Partial-sum(psum, {\color{mypurple}$P$})/Outputs({\color{myred}$O$}). }
    \label{fig:matmul_processing}
    \vspace{-6mm}
\end{figure}
\subsection{Workload Specification} Modern AI inference workloads---from LLM attention blocks to text-to-image
pipelines---are dominated by matrix multiplications and convolutions whose
\emph{operand values and shapes} can vary across layers and requests.
To describe this diversity in a uniform way, we adopt \emph{extended einsum}
notation~\cite{looptree}. Taking matrix multiplication as an example
(Fig.~\ref{fig:matmul_processing}).
\[
{\color{myred}{O}}_{m,n}^{P=M,Q=N} = {\color{myblue}{I}}_{m,k}^{M=M,J=K} \cdot {\color{mygreen}{W}}_{k,n}^{K=K,N=N}
\]
Each tensor is annotated with:

\noindent $\bullet$ \textbf{Superscript}:  an ordered list of \emph{rank names} that define
the logical shape. Ranks are
capitalized, such as ${\color{myblue}{I}}\in\mathbb{R}^{M\times J}, {\color{mygreen}{W}}\in\mathbb{R}^{K\times N},{\color{mypurple}{P}}\in\mathbb{R}^{P\times Q}, {\color{myred}{O}}\in\mathbb{R}^{P\times Q}$. %Superscript is list of rank names, i.e. $[\text{rank}(=\text{shape}),\cdots]$. By convention, rank name is written as capitalized character, and indicates the shape of itself.

\noindent $\bullet$ \textbf{Subscript}: an ordered list of quasi-affine \emph{index
expressions} over rank variables and constants (e.g., $[m,k]$). Rank variables
use the lowercase form of the corresponding rank name. 
% \paragraph{Operands and notation}
  %We use blue/green/purple/red for Input ({\color{myblue}$I$})/ Weight ({\color{mygreen}$W$})/ Partial-sum (psum, {\color{mypurple}$P$})/ Outputs ({\color{myred}$O$}) as shown in \figref{fig:matmul_processing}.% We also add \textbf{Partial Sum (P)} to indicate the intermediate results generated by each PE:

% For standard GEMM, $J=K$ and $(P,Q)=(M,N)$.  We use color coding in this paper: blue/green/purple/red for input({\color{myblue}$I$})/weight({\color{mygreen}$W$})/partial-sum(psum, {\color{mypurple}$P$})/outputs({\color{myred}$O$}).
Partial sums represent intermediate accumulation states that arise when the rank variable of reduction rank (e.g., $k$) is distributed across non-consecutive loop levels in mapping, requiring aggregation of partial results to produce final outputs.
% Partial sums occur when rank variable $k$ of the reduction rank ($J$ and $K$) is partitioned into multiple non-consecutive loop levels in mapping, which generates an accumulation of a subset of $K$ together for each element in outputs. %must be temporarily stored and accumulated into outputs. %Subscript is an quasi-affine expression of rank variables and constants, e.g. $[$expression(rank\_variables, constants),$\cdots]$. \textbf{Rank variable} is the variable in the subscript in a Einsum, which is a small character of the rank name by convention. 
\subsection{Mapping and Layout}
A given extended einsum (\figref{fig:matmul_processing}) can be executed with many legal schedules. We refer to a fine-grained schedule of compute and memory accesses as a \textbf{mapping} which specifies the partitioning, ordering and parallelism of ranks\cite{squareloop, krishna2020data}.%, which specifies (i) rank partitioning, (ii) ordering of partitioned ranks,  (iii) parallelization choices, and (iv) shape parameters\footnote{Mapping is a dataflow (partitioning + ordering + parallelism) with shape information~\cite{krishna2020data}. We use
%\emph{mapping} and \emph{dataflow} interchangeably.}
%(Fig.~\ref{fig:matmul_processing}).
% is a parent term of mapping, which  a category of partitioning, ordering and parallelism \textit{without} detailed shape information, such that dataflow  i.e. a mapping without detailed shapes information. }.

Independently, each tensor can be organized differently in on-chip memories.
We use \textbf{layout} to denote the fine-grained organization of data in
the on-chip buffers~\cite{tong2024FEATHER,squareloop}. %Each logical rank is partitioned into \textbf{layout partitioned ranks}; a layout then specifies the loop order of these ranks. Ranks above \texttt{\#line\_bound} determine the order across memory lines, while ranks below determine the order within a line (Fig.~\ref{fig:matmul_processing}).

% Contemporary reconfigurable accelerator\cite{tong2024FEATHER} requires co-switching both mapping and layout across workloads to utilize all its computation units for workloads of various shapes.

\subsection{Reconfigurable Accelerator}

No single dataflow is globally optimal across the wide shape space of modern
operators~\cite{parashar2019timeloop,squareloop}. This motivates
\textbf{reconfigurable accelerators} that can adapt mappings and layouts at
runtime to sustain high utilization across workloads
\cite{kwon2018maeri,sigma_eq,maestro,CGRA,over_gen,polygraph,DSAGen,tong2024FEATHER}.
FEATHER~\cite{tong2024FEATHER} is a recent practical design that enables
low-cost \emph{co-switching} of dataflow and layout, making reconfigurable
execution deployable.

However, FEATHER has two limitations.

\noindent $\bullet$ \textbf{Pre-known weights offline reordered into ideal layout}. This is because in FEATHER, there is only a single reorder-in-reduction NoC designed to support output layout reordering. For weights, the rigid point-to-point connections from buffers to computation array require weights to be preknown and offline reordered into desired layout. It works for general convolution but is inefficient for recent LLMs when both inputs and weights can arrive or change at runtime. Further, the offline reorganized weights might contain \emph{redundant duplicated data in on-chip buffer}.

\noindent $\bullet$ \textbf{Heavy Control Overhead}. Programming individual switches require dominating configurations overhead at scale. 

To eliminate both, we introduce \textbf{FEATHER+} and its programmer view in \secref{sec:feather_plus}, and Virtual Neuron abstraction and VN-based ISA, MINISA, for FEATHER+ in \secref{sec:method}.
\vspace{-1mm}
\section{FEATHER+ and Programming View}
\vspace{-1mm}
\label{sec:feather_plus}
This section introduces FEATHER+ and programming view. %And it analyzes why Virtual Neuron is the exact level of programming abstraction for FEATHER+.

\vspace{-2mm}
\subsection{FEATHER+ Overview}
\vspace{-2mm}
At its compute, FEATHER+ (Fig.~\ref{fig:FEATHER_program_view}) comprises:

\noindent $\bullet$ \textbf{NEST}: a column-wise independent $AH{\times}AW$ PE array. Each PE holds $2\times AH$ local registers (double buffered to hide loading latency) and performs $AH$-element dot product.

\noindent $\bullet$ \textbf{BIRRD}: a multi-stage reordering-in-reduction network
 to change its target bank during the data reduction.

In its memory, FEATHER+ has:

\noindent $\bullet$ \textbf{Streaming Buffer} stores \textbf{``streaming tensor"}, which is pipelined from top to bottom within each NEST column and reused by all PEs in that column.

\noindent $\bullet$ \textbf{Stationary Buffer} stores \textbf{``stationary tensor"} in local PE registers (green). These register values participate in a single dot product with the streaming tensor as it passes through PE.

\noindent $\bullet$ \textbf{Output Buffer} (OB) stores partial sums that require future update, i.e. temporal reduction. OB is the only multi-bank buffer with individual address generation per bank for supporting flexible output layout reordering.

\subsection{Architectural Refinement of FEATHER into FEATHER+}
\begin{figure}[!t]
    \centering
    % \vspace{-3mm}
    \includegraphics[width=\columnwidth]{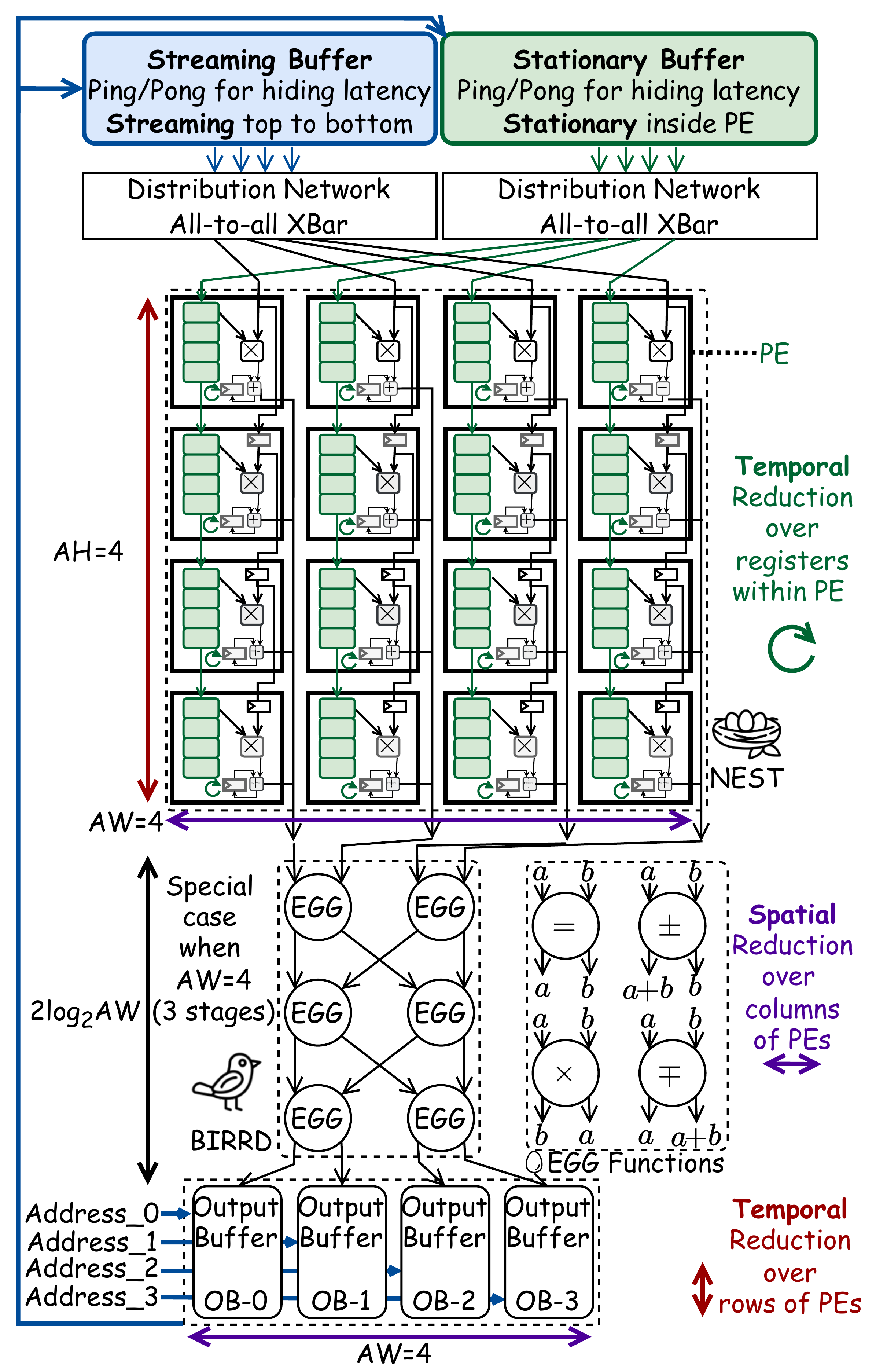}
    \vspace{-6mm}
    \caption{Programmer view of FEATHER+, the reconfigurable accelerator capable of co-switching both mapping and layout.}
    \label{fig:FEATHER_program_view}
    \vspace{-4mm}
\end{figure}

\textbf{FEATHER+} (\figref{fig:FEATHER_program_view}) has three architectural refinement:

\noindent $\bullet$ \textbf{All-to-all Distribution from Buffers to NEST}.
We replace the per-column point-to-point connection from the streaming and
stationary buffers to NEST with two separate all-to-all crossbars.
This allows any resident data to be flexibly multicasted to arbitrary PE columns,
eliminating the need to explicitly materialize duplicated copies of the same
data in on-chip buffers. Such crossbar-style distribution is common
in contemporary AI ASICs\cite{TPUv2, tpuv4i, jouppi2023tpu}, being resource-reasonable. 

\noindent $\bullet$ \textbf{Simplified Streaming Buffer Banking}.
With all-to-all distribution, FEATHER+ no longer requires a multi-bank
streaming-buffer interface to access different rows per bank. 
We regress to single bank with simplified address generation.

\noindent $\bullet$ \textbf{Links from Output Buffer to Stationary Buffer}. This allows current layer’s outputs, used as the next layer’s inputs, to either stay stationary in local registers of PEs or stream through PEs for dot product with the next layer’s weights.

Together, these changes (1) remove the requirement that ``stationary tensor"
must be known ahead of execution to be placed in a preferred layout in stationary buffer, 
(2) free duplicated data in on-chip buffers, and (3) support both inputs and weights to be used as ``stationary tensor". 
FEATHER+ serves as the default hardware backend for this work.

% FEATHER\cite{tong2024FEATHER} might require data duplication in the on-chip buffer to achieve full dataflow/layout flexibility, and require bank-wise address for both streaming buffer (inputs) and output buffers. To avoid data duplication in the on-chip memory and reduce address generation complexity for simplifying control, we augment FEATHER into FEATHER+ with two changes.

% \noindent $\bullet$ \textbf{Distribution Network between Streaming/Stationary Buffer and NEST}: change point-to-point connection between streaming buffer and stationary buffer to each column of PE in NEST into all-to-all crossbars. This ensures on-chip data duplication and removes explicitly loading and storing duplicated data to save off-chip memory accessing latency. We further note that such crossbar is common in today's AI ASICs such as TPU, Trainium etc.

% \noindent $\bullet$ \textbf{Multi-Bank Organization to Single Bank Streaming Buffer}: the introduction of all-to-all distribution network avoid the requirement of multi-bank structure of streaming buffer, hence we simplify it into a single bank.

% Together, these two modifications (i) remove the requirement that weights must be known ahead of time, enabling runtime-arrived dynamic inputs/weights, and (ii) ensure duplication-free data movement across on-chip stationary and streaming buffers. We term this augmented architecture as FEATHER+, which becomes the default reconfigurable accelerator used as backend.

\subsection{Programming View of FEATHER+}

FEATHER+ is a \emph{constrained dataflow accelerator}\cite{constrainted_dataflow_accelerator}. %, enabling flexible reduction patterns among PEs but requires rigid temporal reduction within each PE and temporal reuse of ``streaming tensor" within each column of NEST.
A programmer needs to map workloads into ``\emph{three-level reduction}" with ``\emph{two dataflow choices}" under ``\emph{two constraints}".

\subsubsection{Flexible Mappings}

\paragraph{Three-level Temporal and Spatial Reduction} \hfill

\noindent $\bullet$ \textbf{Temporal reduction within a PE} performed by an $AH$-element dot product between local registers and streamed data over time, which generates a single psum.

\noindent $\bullet$ \textbf{Spatial reduction} over \textbf{a row of PEs via BIRRD}.

\noindent $\bullet$ \textbf{Temporal reduction} over \textbf{a column of PEs in OB}.

Such flexible reduction allows any selected PEs to be grouped together to compute one dot product. 

\paragraph{Two Flexible Mixed Dataflow} \hfill

\noindent $\bullet$ \textbf{Input-Output Stationary (IO-S)}: store inputs in each PE, stream weights, and accumulate psums locally within PE.

\noindent $\bullet$ \textbf{Weight-Output Stationary (WO-S)}: store weights in each PE, stream inputs, and accumulate psums locally within PE.

For MatMul, pick IO-S when $M > N$, otherwise WO-S.
% \subsubsection{Constraints 1: Intra-PE Dot Product} All $AH$ data stationary inside local registers of a single PE must participate into the same dot product, i.e. stationary tensor in each PE should be reduced over time.

% \subsubsection{Constraints 2: Intra-Column Data Reuse} The stationary tensor inside all PEs of the same column must perform dot product with the streaming tensor passing through it. In other words, the streaming tensor should be reused by all PEs of the same column. But all columns in NEST are completely independent.
\subsubsection{Mapping Constraints} 

\paragraph{Constraint 1: Intra-PE Dot Product} All $AH$ data in a PE's local registers must participate in the same dot product, ensuring each PE's stationary tensor is reduced over time. 

\paragraph{Constraint 2: Intra-Column Data Reuse} The streaming tensor is reused across all PEs within the same column to compute dot products with their respective stationary tensors. However, all columns in NEST operate entirely independently.

% \noindent $\bullet$ \textbf{Flexibility: inter-column independence.}

% \subsubsection{Flexibility: Three-level Hierarchical Reduction}

% Intra-Column Data Reuse and Inter-Column Independency

% With VN hidding the dot product behind, the only mapping constraint is \textbf{column-wise \ivn{} reuse}. Specifically, {\ivn{}} moves from top to bottom in the pipeline manner within each PE column. Correct local reduction requires \emph{\wvn} of all PEs in one column to be dot-product with the same \ivn{} stream.

% Further, for each column of PE, inputs move from top to bottom in the pipelined manner. To ensure correct accumulation, weights in the same column must correspond to the same reduction index (e.g. value of $k$ in \figref{fig:matmul_processing} for matrix multiplication).

% Such reconfigurability under the constraints makes FEATHER a constrained dataflow accelerator, i.e. an accelerator capable of supporting subset of arbitrary dataflows to balance resource overhead and performance\cite{constrainted_dataflow_accelerator}.

\subsection{Control Overhead of FEATHER+}

\label{sec:overhead}
While FEATHER+ supports board mapping/layout choices, \emph{realizing} this
flexibility via switch- and cycle-level micro-control can incur substantial
overheads. Programs must specify control for BIRRD and buffer address generation for each cycle.
This has two direct consequences.
First, control state consumes on-chip storage that could otherwise be used for
activations and weights, limiting the largest feasible tile size and reducing
arithmetic intensity.
Second, fetching long instruction traces increases off-chip instruction
traffic, which is especially harmful for memory-bound layers.
As accelerator arrays scale, this control
pressure becomes a first-order limiter of end-to-end throughput. When adopting micro-control to configure each individual switches for FEATHER+, significant memory overhead of instructions fetching would dominate at larger scale, as shown in \tabref{tab:compute_stall_instruction}. Such control overhead motivates MINISA -- a low-cost abstraction of hardware flexibility.

\begin{table}[!htp]\centering
\vspace{-2mm}
\caption{Explicit stall of fetching instructions for matrix multiplication of $\sum_{k}{\color{myblue}{I}}_{m,k}^{M=65536,J=40} \cdot {\color{mygreen}{W}}_{k,n}^{K=40,N=88}$}\label{tab:compute_stall_instruction}
\vspace{-2mm}
\resizebox{\columnwidth}{!}{
\begin{tabular}{lccccccc}\hline
FEATHER &$4{\times}4$ &$8{\times}8$ &$4{\times}64$  &$16{\times}16$ &$8{\times}128$ &$16{\times}256$ \\\hline
stall &0 &0 &75.3\% &65.2\% &90.4\% &96.9\% \\
\hline
\end{tabular}}
\vspace{-4mm}
\end{table}

% To unleash the flexibility of reconfigurable accelerator like FEATHER to achieve (dataflow, layout) co-switching at runtime requires changing configurations per cycle, introducing significant control overhead. For a single convolution layer e.g. the first layer in ResNet50, which requires reconfigurable overhead of $how much$, taking $how much$\% of the entire off-chip memory access, introducing (1) significant on-chip buffer capacity requirement for storing control, which could have been used to store the actual data instead, and (2) higher off-chip memory traffic, worsen memory-bounded workloads.

% \subsection{Insight: Coarse-grained Reconfigurability at Dot Products} 

% To reduce the significant control overhead while not sacrificing overall flexibility provided by the hardware, MINISA bases on the key insight that FEATHER's always perform computation at the granularity of PE, i.e. a single dot product of size $AH$. Therefore, we could reconfigure the mapping and layout at the granularity of $AH$-sized dot product instead of individual PE or switches in the hardware to retain its flexible three-level reduction while reducing the redundant control overhead.

\section{MINISA}
\label{sec:method}

This section introduces MINISA, a minimal instruction set that configures FEATHER+ at the Virtual Neuron (VN) granularity. We detail the VN abstraction, its mapping constraints under VN, the corresponding ISAs, and the execution model.
% \subsection{Overview of MINISA}
% MINISA is built on a simple observation: FEATHER+'s minimal compute unit
% is a PE that performs an $AH$-element dot product. Rather than configuring
% switch-level micro-operations, MINISA treats this $AH$-element dot product
% as the atomic unit of control. This abstraction preserves FEATHER+'s built-in
% flexibility while compressing the prohibitively large switch-level configuration space into a small set of symbolic VN-based layout and mapping abstractions.

\vspace{-2mm}
\subsection{Key Idea: VN (AH-element Dot Product) as Abstraction}
MINISA is inspired by a simple architectural observation:

\textit{FEATHER+ always computes at the granularity of an $AH$-element dot product in its PE, and supports flexible reduction over the the dot-product results from any PEs.}

We therefore define a \textbf{Virtual Neuron}\footnote{we borrow this term from MAERI\cite{kwon2018maeri} which was a flexible dataflow accelerator that defined VN to be a flexible sized dot product engine. But here we use it as the abstraction of a fix-size dot product performed by a PE.} as the smallest software operand fragment that matches this hardware atom, i.e. $AH$-element dot product. This leads to a key insight:

\emph{Programming at VN granularity is the coarsest control that still preserves
inter-PE mapping flexibility and the finest control that avoids unnecessary
switch-level overhead.}

Concretely, MINISA re-expresses FEATHER+'s legal mapping and layout space
as the granularity of VN. This factorization retains FEATHER+'s built-in reconfigurability while
dramatically compressing instruction footprint, enabling larger on-chip tiles and higher end-to-end performance.

% \subsubsection{Justification of Minimum} A coarser-grained abstraction than current VN would ignore inter-PE mapping flexibility, while a fine-grained abstraction introduces unnecessary costs without improving flexibility, hence programming at the level of VN would be the exact level to minimize control overhead without losing hardware built-in flexibility.

\subsection{MINISA Abstraction -- Virtual Neuron (VN)}

\subsubsection{Definition}
A \emph{Virtual Neuron} is the minimal hardware dot-product atom, which is also the largest atomic unit in software (compilation) without losing flexibility in hardware. For an FEATHER+ with $AH\!\times\! AW$ NEST, each PE performs an $AH$-element dot product, hence the VN size should be smaller than or equal to $AH$. We use VN size as $AH$ in the following of this paper for clear idea demonstration.

\subsubsection{Operand-specific VNs}
Different operands have their own VNs, including inputs (\ivn{}), weights (\wvn{}), partial sums (\pvn{}) and outputs (\ovn). Each VN is obtained by grouping consecutive AH elements along the reduction dimension, i.e. $J$ of inputs, $K$ of weights and $Q$ of partial sum and outputs (which becomes $J$ of input for the layer following it). For example, weight VNs are indexed by \wvn{}$(r,c)$, where $r\!\in\![0,\lceil K//AH\rceil-1]$ and $c\!\in\![0,N-1]$ denote the row and column indices. In this way, \ivn{} and \wvn{} exactly denote a group of elements from inputs and weights to be \textbf{consumed by local dot product in one PE}, respectively. Each PE's dot product yields a single element in either \pvn{} or \ovn, depending on whether further reduction is required.

\subsubsection{FEATHER+ Mapping Constraints under VNs}

The VN abstraction hides Constraint 1, leaving Constraint 2 (Intra-Column \ivn{} Reuse) as the only restriction. \ivn{} pipelines top-to-bottom within each NEST column, requiring all PEs in a column to compute dot products between their \emph{\wvn} and this shared \ivn{} stream for correct local reduction.

\subsection{MINISA ISA Overview}

\begin{table}[h]
  \centering
  \caption{Overview of MINISA instructions.}
  \vspace{-2mm}
  \label{tab:minisa-overview}
  \resizebox{\columnwidth}{!}{
  \begin{tabular}{ll}
    \hline
    ISA name & ISA meaning \\
    \hline
    \texttt{SetIVNLayout} &
    Configure data layout for \ivn{} in on-chip buffer. \\[0.2em]
    \texttt{SetWVNLayout} &
    Configure data layout for \wvn{} in on-chip buffer. \\  \hdashline
    \texttt{SetOVNLayout} &
    \makecell[l]{Configure the \emph{output} buffer layout for output VNs and \\ clear (initialize) the on-chip output buffer for accumulation.} \\ \hdashline 
    \texttt{ExecuteMapping} &
    \makecell[l]{Specify the VN-level mapping parameters and trigger execution of \\ a
    single compute tile on the PE array under the current VN layouts.} \\ \hdashline
    \texttt{ExecuteStreaming} & \makecell[l]{Specify how VNs are streamed in each column \\ and swap Dataflow between IO-S and WO-S.} \\ 
    \hdashline
    \texttt{Load} & Load data from off-chip memory into
    streaming / stationary buffer.  \\  
    % \texttt{LoadWVN} & Load \wvn{} from off-chip memory into
    % streaming / stationary buffer.  \\ 
    \texttt{Write} & write data from streaming / stationary buffer to off-chip memory.  \\ 
    \texttt{Activation} & Perform activation function on data in streaming / stationary buffer. \\ 
    \hline
  \end{tabular}}
  \vspace{-4mm}
\end{table}

\subsubsection{ISA} MINISA defines four ISAs (\tabref{tab:minisa-overview}) to reconfigure mapping and data layout in on-chip buffer at VN granularity, and supporting ISAs for data loading/writing and activation.

\noindent $\bullet$ \texttt{SetIVNLayout} / \texttt{SetWVNLayout}: configure the on-chip streaming and stationary buffer layouts for Virtual Neurons of Input (\ivn) and Weight (\wvn{}).

\noindent $\bullet$ \texttt{SetOVNLayout}: Configures the layout for Output Virtual Neurons (\ovn) in the output buffer, which becomes the layout of input for next successive operation.

\noindent $\bullet$ \texttt{ExecuteMapping}: Configures mapping of stationary tensor (\wvn{} or \ivn{}) to PEs in one compute tile.

\subsubsection{ISA Bitwidth Analysis} 
ISA parameter bitwidths are designed to support the maximum ratio between on-chip buffer capacities and architectural dimensions. Specifically, this applies to the ratio of stationary or streaming buffer depths ($D\!=\!D_{sta}\!=\!D_{str}$) to the NEST width or height ($AW$, $AH$). Workload parameters can be configured to any value within these bitwidth limits; if the actual element index exceeds the workload shape, it is automatically zero-padded.

\begin{figure}[!h]
  \vspace{-3mm}
  \centering
  \includegraphics[width=\columnwidth]{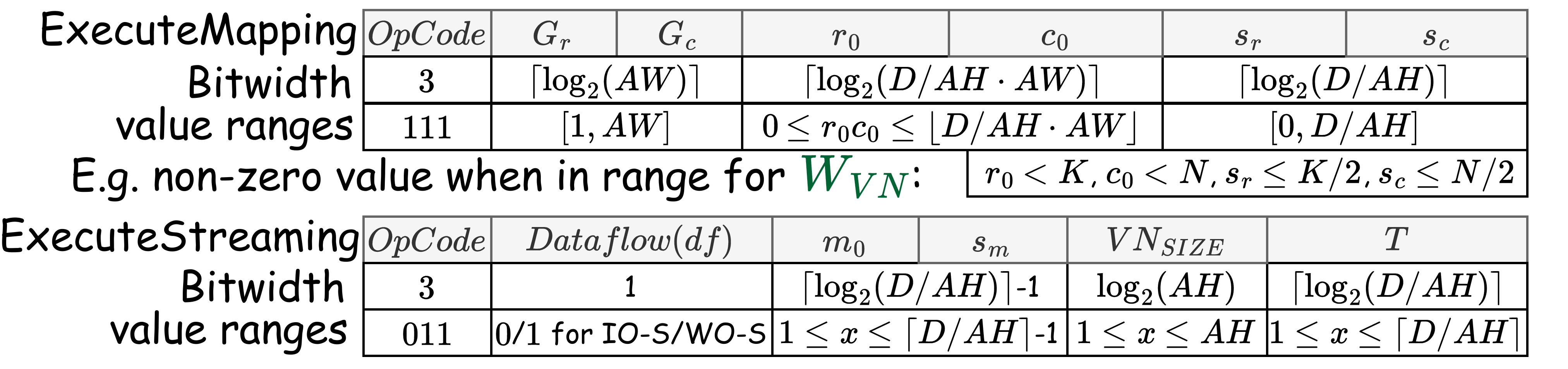}
  \vspace{-7mm}
  \caption{\texttt{Execute*} Field definition for $AH\!\times\! AW$ NEST.}
  \label{fig:minisa_specification}
  \vspace{-5mm}
\end{figure}

\subsection{ExecuteMapping: Flexible Mapping for NEST}
\label{subsec:minisa-nest}

\subsubsection{Definition}
For an $AH \!\times\! AW$ PE array, PE $(a_h,a_w)$ denotes the PE at
row $a_h\!\in\![0,AH)$ and column $a_w\!\in\![0,AW)$. Each PE could holds a single \wvn{}. Therefore, our objective is to specify which \wvn$(r,c)$ is stored in each PE.
Using WO-S dataflow as example, MINISA realizes all legal mappings using six parameters $\theta_{EM} \!=\! (r_0,c_0,G_r,G_c,s_r,s_c)$.
\begin{equation}
  r\!=\! r_0
       + \left\lfloor \frac{a_w}{G_r} \right\rfloor
        \quad
  c\!=\! c_0
       + s_r\,a_h
       + s_c\,\bigl(a_w \bmod G_c\bigr)
\end{equation}

where, ($r_0, G_r$) control placement of \wvn{}s within each column, preserving the architectural constraint that all PEs in one PE column share the same \wvn{} \emph{``row index ($r$)"}.  

\noindent $\bullet$ $r_0$ denotes the starting row index of the \wvn{} in NEST.

\noindent $\bullet$ $G_r$ specifies how many consecutive PE columns share the same \wvn{} row index ($r$) before incrementing to next row index, and hence is bounded by the number of columns ($AW$).

($c_0,s_r,G_c,s_c$) control \wvn{}s across PE columns.

\noindent $\bullet$ $c_0$ selects the starting \wvn{} column index ($c$).

\noindent $\bullet$ $G_c$ defines replication period for $c$ of \wvn{} across columns.

\noindent $\bullet$ $s_c$ determines the spacing in column index ($c$) of \wvn{} among distinct PE-column patterns within one period. 

\noindent $\bullet$ $s_r$ is the temporal stride across PE rows, determining how \wvn{} column indices grow within a PE column. 

Each buffer could holds at most $D\!/\!AH\!\cdot\!AW$ VNs, such that $r_0\cdot c_0$ peaks at $D\!/\!AH\!\cdot\!AW$. Stride peaks at half. Any \wvn$(r,c)$ falling outside the tensor bounds is implicitly zero-padded.

\subsubsection{Case Studies}
\begin{figure}[!t]
  \centering
  \includegraphics[width=\columnwidth]{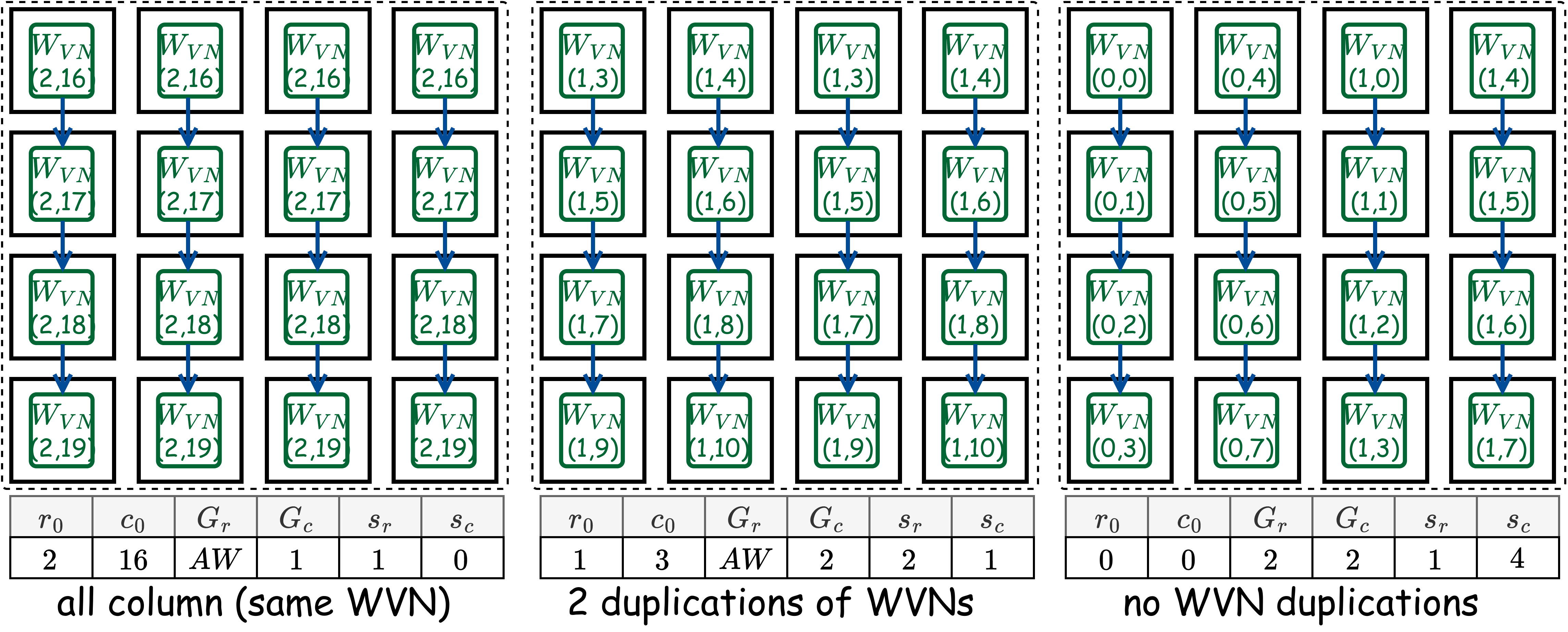}
  \vspace{-7mm}
  \caption{ExecuteMapping examples for $4\!\times\!4$ NEST.}
  \label{fig:example_of_weights_VN_mapping}
  \vspace{-3mm}
\end{figure}

Because columns are independent, \figref{fig:example_of_weights_VN_mapping} shows three ExecuteMapping choices: (1) replicate the same WVNs across all columns, enabling all IVN chunks to be processed in parallel across every column. (2) partition the columns into two replicated groups, allowing the IVN stream to be divided into two independent substreams that are processed in parallel. and (3) assign each column a different set of WVNs, so each column processes an independent IVN stream.

\begin{table}[t]
\centering
\small
\caption{Unified 3-bit permutation encoding across VN layouts.
Each column applies the same \texttt{order\_id} value to the operand-specific
three-rank set remaining after the VN constraint
($K_{L0}\!=\!AH$, $J_{L0}\!=\!AH$, $Q_{L0}\!=\!AH$).}
\vspace{-2mm}
\label{tab:vn-orderid-all}
\resizebox{\columnwidth}{!}{
\begin{tabular}{c|c|c|c}
\hline
\textbf{\texttt{order}}
& \textbf{\wvn{} order}
& \textbf{\ivn{} order}
& \textbf{\ovn{} order} \\
\hline
000
& $k_{L1} \rightarrow n_{L0} \rightarrow n_{L1}$
& $j_{L1} \rightarrow m_{L0} \rightarrow m_{L1}$
& $p_{L1} \rightarrow p_{L0} \rightarrow q_{L1}$ \\
001
& $k_{L1} \rightarrow n_{L1} \rightarrow n_{L0}$
& $j_{L1} \rightarrow m_{L1} \rightarrow m_{L0}$
& $p_{L1} \rightarrow q_{L1} \rightarrow p_{L0}$ \\
010
& $n_{L0} \rightarrow k_{L1} \rightarrow n_{L1}$
& $m_{L0} \rightarrow j_{L1} \rightarrow m_{L1}$
& $p_{L0} \rightarrow p_{L1} \rightarrow q_{L1}$ \\
011
& $n_{L0} \rightarrow n_{L1} \rightarrow k_{L1}$
& $m_{L0} \rightarrow m_{L1} \rightarrow j_{L1}$
& $p_{L0} \rightarrow q_{L1} \rightarrow p_{L1}$ \\
100
& $n_{L1} \rightarrow k_{L1} \rightarrow n_{L0}$
& $m_{L1} \rightarrow j_{L1} \rightarrow m_{L0}$
& $q_{L1} \rightarrow p_{L1} \rightarrow p_{L0}$ \\
101
& $n_{L1} \rightarrow n_{L0} \rightarrow k_{L1}$
& $m_{L1} \rightarrow m_{L0} \rightarrow j_{L1}$
& $q_{L1} \rightarrow p_{L0} \rightarrow p_{L1}$ \\
110/111
& reserved
& reserved
& reserved \\
\hline
\end{tabular}}
\vspace{-4mm}
\end{table}

\begin{figure}[!t]
  \centering
  \includegraphics[width=\columnwidth]{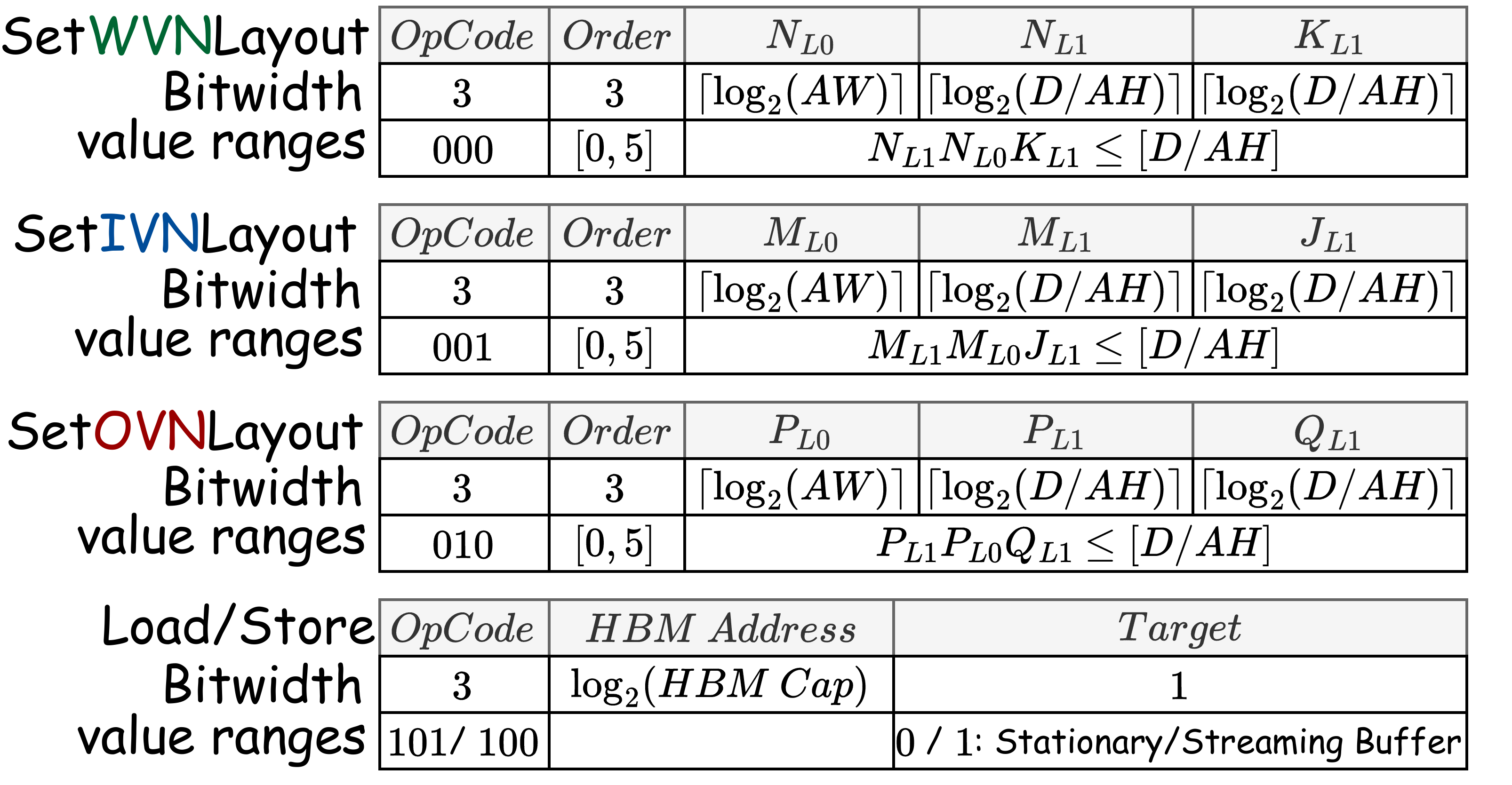}
  \vspace{-5mm}
  \caption{MINISA specifications for layout, load/store, and dataflow. $D$ refers to the depth of stationary / streaming buffer.}
  \label{fig:minisa_iwovn}
  \vspace{-4mm}
\end{figure}

\subsection{ExecuteStreaming}
\subsubsection{Definition}
Under WO-S, streamed \ivn{}s enter from the top of each PE column and are reused across consecutive cycles as they propagate downward. Therefore, we shall specify which \ivn$(m,j)$ is injected into each PE column over time.%, where $m$ is the streamed row index and $j$ is reduction rank variable.

To minimize ISA overhead, \texttt{ExecuteStreaming} reuses parameters $\theta_{EM}$
defined by the paired \texttt{ExecuteMapping}, and introduces give additional parameters $\theta_{ES} = (m_0, s_m, T, VN_{size}, df)$.

\noindent $\bullet$ $m_0$ is the starting streamed-row index. 

\noindent $\bullet$ $s_m$ is the temporal stride of streamed \ivn{}. 

\noindent $\bullet$ $T$ is the number of \ivn{} injected into each PE column.

\noindent $\bullet$ $df$ changes the dataflow choice (\texttt{IO-S} or \texttt{WO-S}).

\noindent $\bullet$ $VN_{size}$ defines the size of Virtual Neuron.

All fields encode ``value-1" omitting zero to reduce bitwidth.

Given the paired \texttt{ExecuteMapping}, the \ivn{}($m$, $j$) injected into PE column $a_w$ at temporal step $t\!\in\![0,T)$ is
\begin{equation*}
j \!=\! r_0 + \left\lfloor \frac{a_w}{G_r} \right\rfloor,
\quad 
m \!=\! m_0
    + s_m\,t
    + \left\lfloor \frac{a_w \bmod G_r}{G_c} \right\rfloor.
\end{equation*}

\subsubsection{Case Study}

As an example, consider an $AH \!\times\! 4$ PE array with the paired
\texttt{ExecuteMapping} parameters chosen as
$(r_0,G_r,G_c)=(0,2,1)$, and \texttt{ExecuteStreaming} set to
$(m_0,s_m,T,df)=(0,3,3,1)$. PE columns $0$ and $1$
belong to reduction group $j=0$, while PE columns $2$ and $3$ belong to
reduction group $j=1$. Within each two-column group, the two columns split and process overall \ivn{} in parallel. Therefore, over three injection cycles, the \ivn{}s entering the top of the four PE columns are:

{\small \[
\begin{array}{c|c|c|c|c}
2  & \text{\ivn}(6,0) \downarrow& \text{\ivn}(7,0) \downarrow& \text{\ivn}(6,1) \downarrow& \text{\ivn}(7,1) \downarrow\\ 
1 & \text{\ivn}(3,0)  \downarrow& \text{\ivn}(4,0) \downarrow & \text{\ivn}(3,1)  \downarrow& \text{\ivn}(4,1)  \downarrow\\
0& \text{\ivn}(0,0) \downarrow  & \text{\ivn}(1,0) \downarrow  & \text{\ivn}(0,1)  \downarrow & \text{\ivn}(1,1) \downarrow  \\
\hline
\text{cycle} & \text{PE col }0& \text{PE col }1 & \text{PE col }2 & \text{PE col }3 \\
\end{array}
\]}
In the array, each column corresponds to one PE column in hardware, while each row denotes the \ivn{} injected at the top of that PE column in one cycle. After injection, each \ivn{} propagates downward through the corresponding PE column over consecutive cycles, so the same streamed tensor (\ivn{} under WO-S) is temporally reused by all PEs in that column.

\subsection{Set*VNLayout: Flexible Layout for On-chip Buffers}
\subsubsection{Key Idea: 4-level Partitioned Tensor in AW-width Buffer} 
The three layout instructions (\texttt{SetIVNLayout}, \texttt{SetWVNLayout}, and \texttt{SetOVNLayout}) define how a logical 2-rank tensor is placed into a physical $D\!\times\!AW$ on-chip buffer. Each instruction targets different operand types (\ivn{}/\wvn{}/\pvn{} and \ovn{}). Under WO-S, \ivn{} is mapped to the streaming buffer and \wvn{} to the stationary buffer. Under IO-S, these roles are swapped.

At a high level, layout specification has three steps: (1) choose rank partitioning factors, (2) choose an ordering of the post-partitioned ranks, and (3) fold the resulting VN sequence into the $D\!\times\!AW$ buffer.

\subsubsection{Partitioning Factors}
Using the weight matrix $(K,N)$ as an example, we first express layout at element granularity.
Each rank is split into two levels:
\[
K \!=\! K_{L1}K_{L0},\qquad N \!=\! N_{L1}N_{L0},
\]
\[
k \!=\! k_{L1}K_{L0} + k_{L0},\qquad n \!=\! n_{L1}N_{L0} + n_{L0}.
\]
This yields four loop ranks $\{k_{L0},k_{L1},n_{L0},n_{L1}\}$, which can represent both contiguous and strided layouts.

VNs always group consecutive elements along the reduction rank (i.e., $K/J/Q$ for \wvn{}/\ivn{}/\ovn{} in \figref{fig:layout_illustration_gran}), therefore MINISA fixs the innermost reduction-level factor at the size of VN, e.g., enforces $K_{L0}$ as VN size for \wvn{}. VN size can be any value up to $AH$. This paper uses $AH$ for illustration. 

Further, all elements of a single VN are accessed serially across multiple cycles (rather than in a single-cycle concurrent access), all elements of that VN are placed in contiguous buffer rows at a fixed buffer-column index. 

\subsubsection{Ordering}
The VN abstracts away the level-0 partitioning factor of the reduction rank, the remaining freedom is only the order of the three other remaining ranks, e.g., $\{K_{L1},\;N_{L0},\;N_{L1}\}$ for \wvn{}. This reduced loop space is still complete for VN-granularity layouts.
Under such rank partitioning, we identify weight VNs by \wvn{}$(r,c)$, where
\[
  r \!=\! k_{L1},\qquad
  c \!=\! n_{L1}\cdot N_{L0} + n_{L0}.
\]

\paragraph{Addressing Generation}
We first flatten all VNs according to the chosen rank order to obtain a 1D index $L$, and then map $L$ to buffer coordinates by row-major placement over the $D\times AW$ buffer. For \wvn{} with ranks $\{K_{L1},N_{L0},N_{L1}\}$,
\[
\mathbf{R} \!=\! [K_{L1},\;N_{L0},\;N_{L1}],\qquad
\mathbf{RV} \!=\! [k_{L1},\;n_{L0},\;n_{L1}],
\]
where $\mathbf{R}$ denotes the ordered ranks and $\mathbf{RV}$ their associated rank variables. Let $P\!=\!(p_0,p_1,p_2)$ be a permutation of ${0,1,2}$ defining the loop order from outermost to innermost. The resulting flattened VN index is
\[
  L
  \!=\! RV_{p_0}\cdot R_{p_1}R_{p_2}
  + RV_{p_1}\cdot R_{p_2}
  + RV_{p_2}
\]
and the physical buffer address is
\[
  addr_{row}\!=\!\left\lfloor \frac{L}{AW}\right\rfloor,\qquad
  addr_{col}\!=\!L \bmod AW.
\]

\subsubsection{Bitwidth Analysis}
\paragraph{Order} Arbitrary ordering of three partitioned ranks offers $3!\!=\!6$ legal permutation choices, requiring
$\lceil \log_2 6\rceil\!=\!3$ bits. Therefore, \emph{order} is encoded with 3 bits. The operand-specific mapping of each code is listed in \tabref{tab:vn-orderid-all}.

\paragraph{Partitioning Factors} Partition factors determine size of  tensors which must fit in overall buffer capacity. For \wvn{}, the number of VNs that could be stored by a $D\times AW$ buffer is bounded by
$K_{L1}\cdot N_{L1}\cdot N_{L0} \le \left\lfloor \frac{D}{AH} \right\rfloor \cdot AW$. This restricts the overall bitwidth.
% Equivalently, in element form, $K_{L0}K_{L1}N_{L1}\le\left\lfloor \frac{D}{AH} \right\rfloor$ with $K_{L0}\!=\!AH$.
We also cap level-0 non-reduction factors (e.g., $N_{L0}\!\le\!AW$), since larger values are performance-equivalent to existing value within $AW$.

\subsubsection{Case study} 
\begin{figure}[!t]
    \centering
    % \vspace{-3mm}
    \includegraphics[width=0.98\columnwidth]{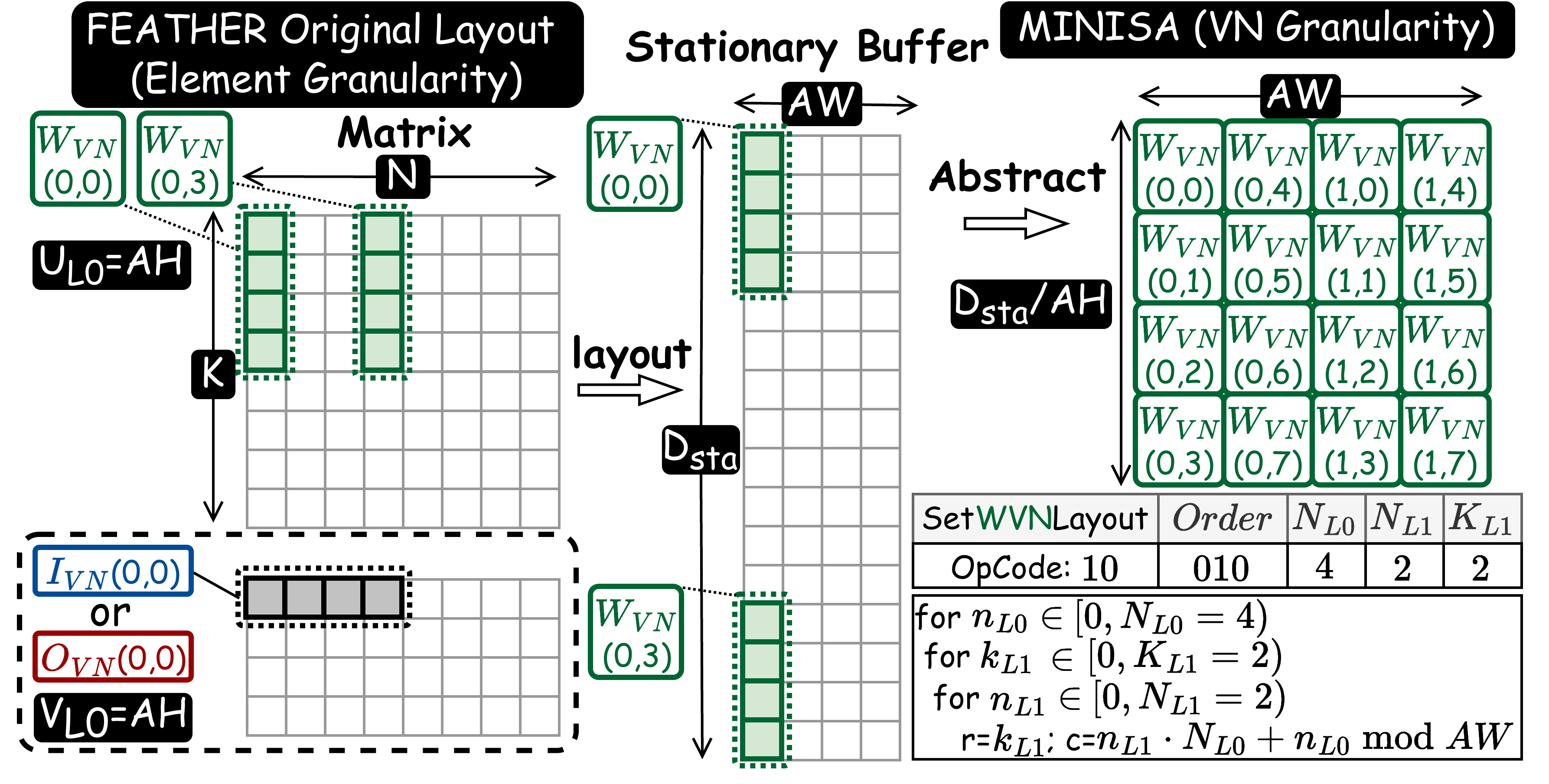}
    \vspace{-3mm}
  \caption{MINISA for layout Specification and \texttt{SetWVNLayout} illustration ($K\!=\!8, N\!=\!8, AH\!=\!AW\!=\!4
  \Rightarrow K_{L0}\!=\!AH$). Takeaway: MINISA organizes layout at VN granularity.}
    \label{fig:layout_illustration_gran}
    \vspace{-3mm}
\end{figure}

\figref{fig:layout_illustration_gran} uses loop order $n_{L0}\!\rightarrow\!k_{L1}\!\rightarrow\!n_{L1}$ (with $N_{L0}\!=\!4$, $K_{L1}\!=\!2$, $N_{L1}\!=\!2$). For $n_{L0}\!=\!0$, the first buffer row places
$\swvn{}(0,\,0),\swvn{}(0,\,4),\swvn{}(1,\,0),\swvn{}(1,\,4)$ across the $AW\!=\!4$ columns. The same pattern repeats for $n_{L0}\!=\!1,2,3$.
This case study makes explicit that \texttt{SetWVNLayout} encodes the full legal layout spaces with a compact VN-level descriptor (partition factors + order).

\begin{figure}[!t]
    \centering
    % \vspace{-3mm}
    \includegraphics[width=0.98\columnwidth]{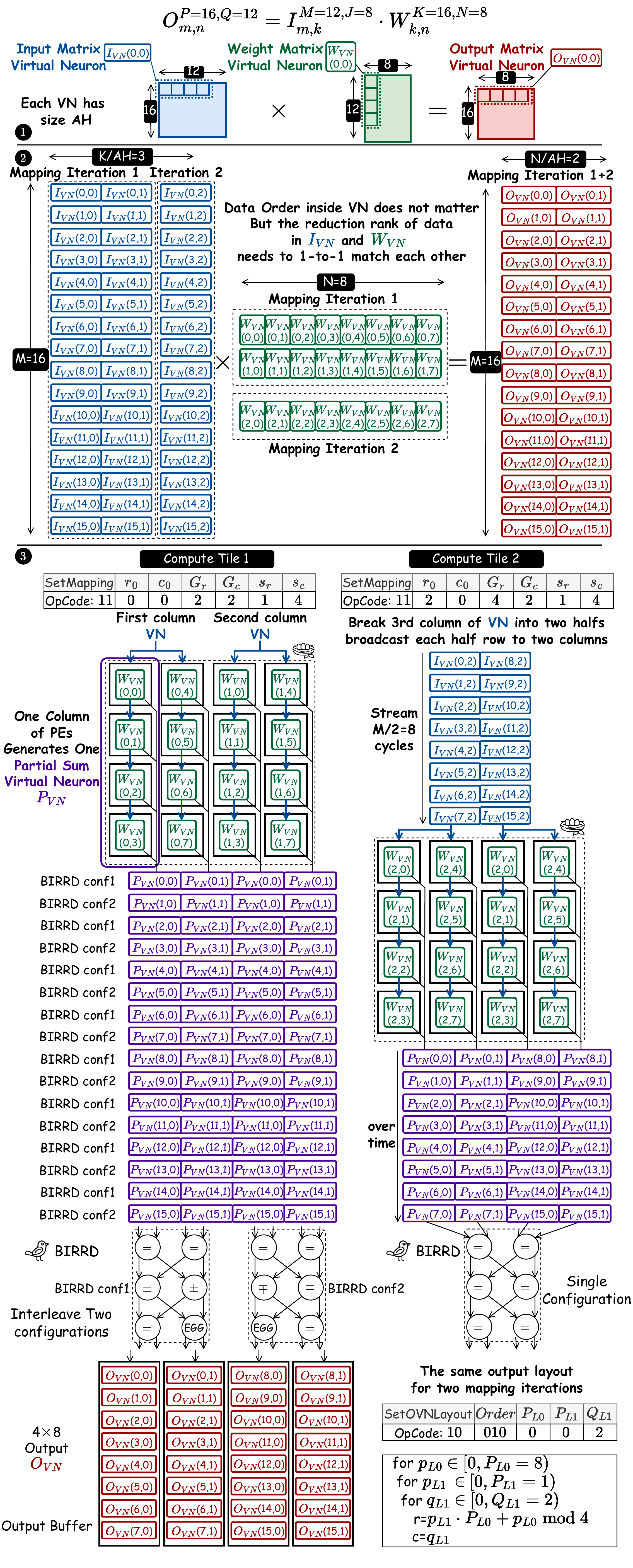}
    \vspace{-4mm}
  \caption{Mapping a matrix multiplication to FEATHER+ (AH$\!\times\!$AW$\!=\!$4$\!\times\!$4 NEST) as 2 compute tiles under MINISA.}
    \label{fig:workload_case_study}
    \vspace{-8mm}
\end{figure}

\subsection{Execution Model}

\subsubsection{Roles of MINISA ISAs}
MINISA cleanly separates \emph{configuration}, \emph{data movement}, and \emph{compute triggering}.

\noindent $\bullet$ \textbf{Configuration-only ISAs:}
\texttt{Swap}, \texttt{SetIVNLayout}, and \texttt{SetWVNLayout} only program internal control/state registers. They do not move data and do not launch computation.

\noindent $\bullet$ \textbf{Memory-movement ISAs:}
\texttt{Load} and \texttt{Write} trigger data transfers between off-chip memory and streaming/stationary buffers.
\texttt{SetOVNLayout} additionally manages output-buffer lifecycle at tile boundaries: it initializes the output tile before accumulation and commits the finished tile from output buffer to the next operand buffer according to dataflow (IO-S: streaming, WO-S: stationary). 

\noindent $\bullet$ \textbf{Compute-trigger ISA:}
FEATHER+ only triggers on-chip activity when reciving \texttt{E.Streaming} and \texttt{E.Mapping}. Each issue loads stationary tensors into NEST from on-chip buffers, and computes on one tile under current layouts.

\noindent $\bullet$ \textbf{Sub-tiled execution:}
Layout configurations are reused across multiple pairs of (\texttt{ExecuteMapping}, \texttt{ExecuteStreaming}) issues, enabling a sequence of sub-tiles to contribute on the same outputs.

\subsubsection{Execution Model}
The canonical trace for one layer is:
\[
\texttt{Set*VNLayout}
\;\rightarrow\;
\{\texttt{E.Mapping/E.Streaming}\}^{\!\times\! T}
\]
where \(T\) is the number of compute tiles needed to consume the currently loaded \ivn{}/\wvn{} for that layer.
The layout instructions first define operand placement and initialize the output tile, and the following \texttt{ExecuteMapping} sequence processes all tiles with unique mapping.

For consecutive layers, the output of layer~\(i\) becomes the input of layer~\(i{+}1\).
Therefore, \texttt{SetIVNLayout} is issued once for the first layer, and the \texttt{SetOVNLayout} of layer~\(i\) is reused as the \texttt{SetIVNLayout} of layer~\(i{+}1\), which could be optionally skipped. 
Each new layer still issues its own \texttt{SetWVNLayout}, \texttt{SetOVNLayout}, and a batch of \texttt{ExecuteMapping} instructions.

\subsubsection{Walk-through Case Study}
\figref{fig:workload_case_study} illustrates this execution semantics for matrix multiplication.
Assume \texttt{SetIVNLayout}, \texttt{SetWVNLayout} and \texttt{Load} have already established layouts and loaded required VNs.
The on-chip tensors are then consumed by two successive \texttt{ExecuteMapping/Streaming} instructions (\ding{184}) on the same \(AH\!\times\!AW\) NEST.
Both tiles share the same \texttt{SetOVNLayout} and accumulate \pvn{}s into the same final \ovn{}. 
Because layout of \ovn{} remain invariant across the two mappings, \pvn{}s from both tiles are guaranteed to accumulate into consistent locations, completing the full \ovn{} without extra output re-layout or intermediate data movement.

\section{Compilation Pipeline}
\label{sec:compilation}

\begin{figure}[!t]
  \centering
  \includegraphics[width=\columnwidth]{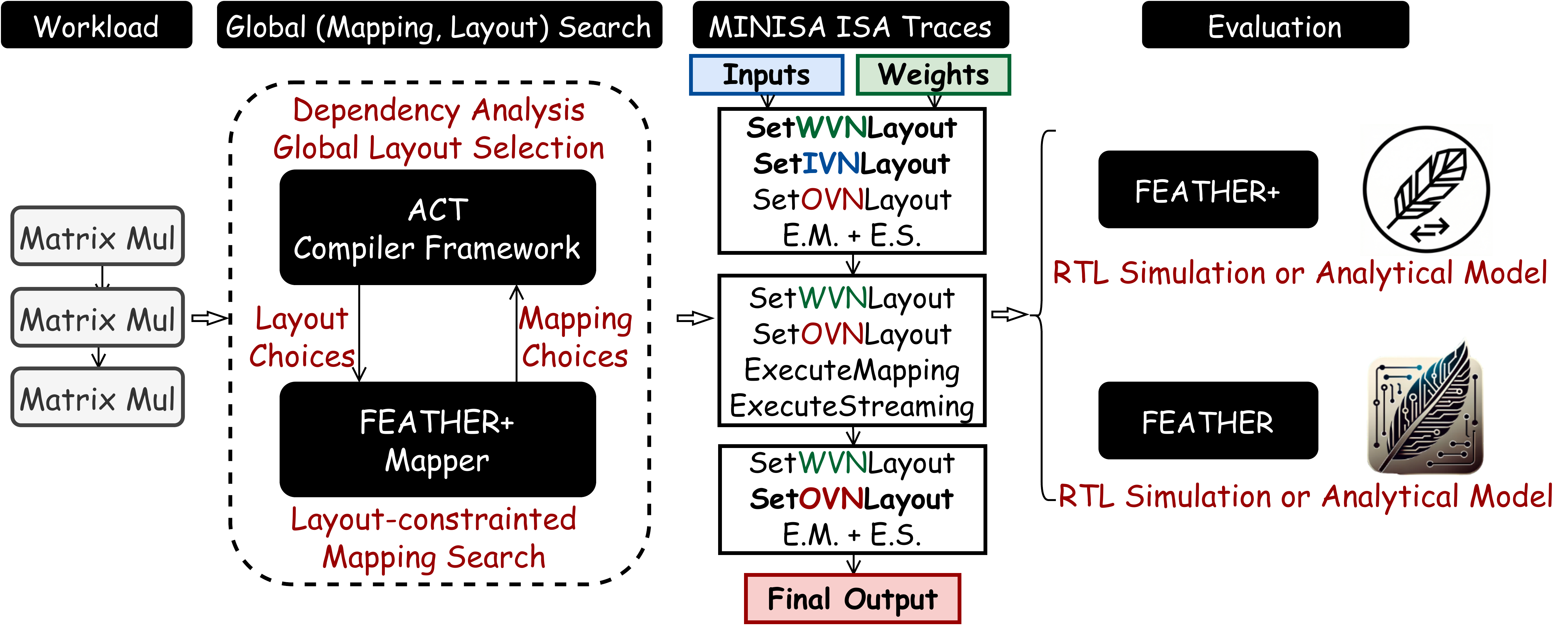}
  \vspace{-6mm}
  \caption{Compilation flow for FEATHER+.}
  \label{fig:compilation_flow}
  \vspace{-3mm}
\end{figure}

\subsection{Overview}
For a workload with a chain of matrix multiplication, we provide an analytical (mapping, layout) co-search framework, FEATHER+ Mapper. Chosen (dataflow, layout) is deterministically translated into a trace of MINISA ISAs, then lowered into FEATHER+ ``micro-instructions".

For a workload represented as a chain (or DAG) of layers including matrix multiplications and activations (e.g., softmax), we integrate FEATHER+ Mapper as a layout-constrained mapping search framework into ACT Ecosystem\cite{act_arxiv, jain2025taidl}, as shown in \figref{fig:compilation_flow}. Specifically, ACT first performs graph-level analyses and identifies layout-flexible regions—subgraphs where tensor layouts may be changed while maintaining correctness, subject to any required layout constraints at region boundaries. For each layer in such a region, ACT invokes FEATHER+ mapper to perform layout-constrained mapping search. The selected per-layer mapping choices are sent back to ACT to finalize the global (mapping, layout) choice with the lowest latency. The final choices are lowered into MINISA ISA traces, which are executed in FEATHER / FEATHER+ RTL simulation or analytical simulation model.

\subsection{Overall Flow of the FEATHER+ Mapper}
\label{subsec:mapper_overall_flow}

\begin{figure}[!t]
  \centering
  \includegraphics[width=\columnwidth]{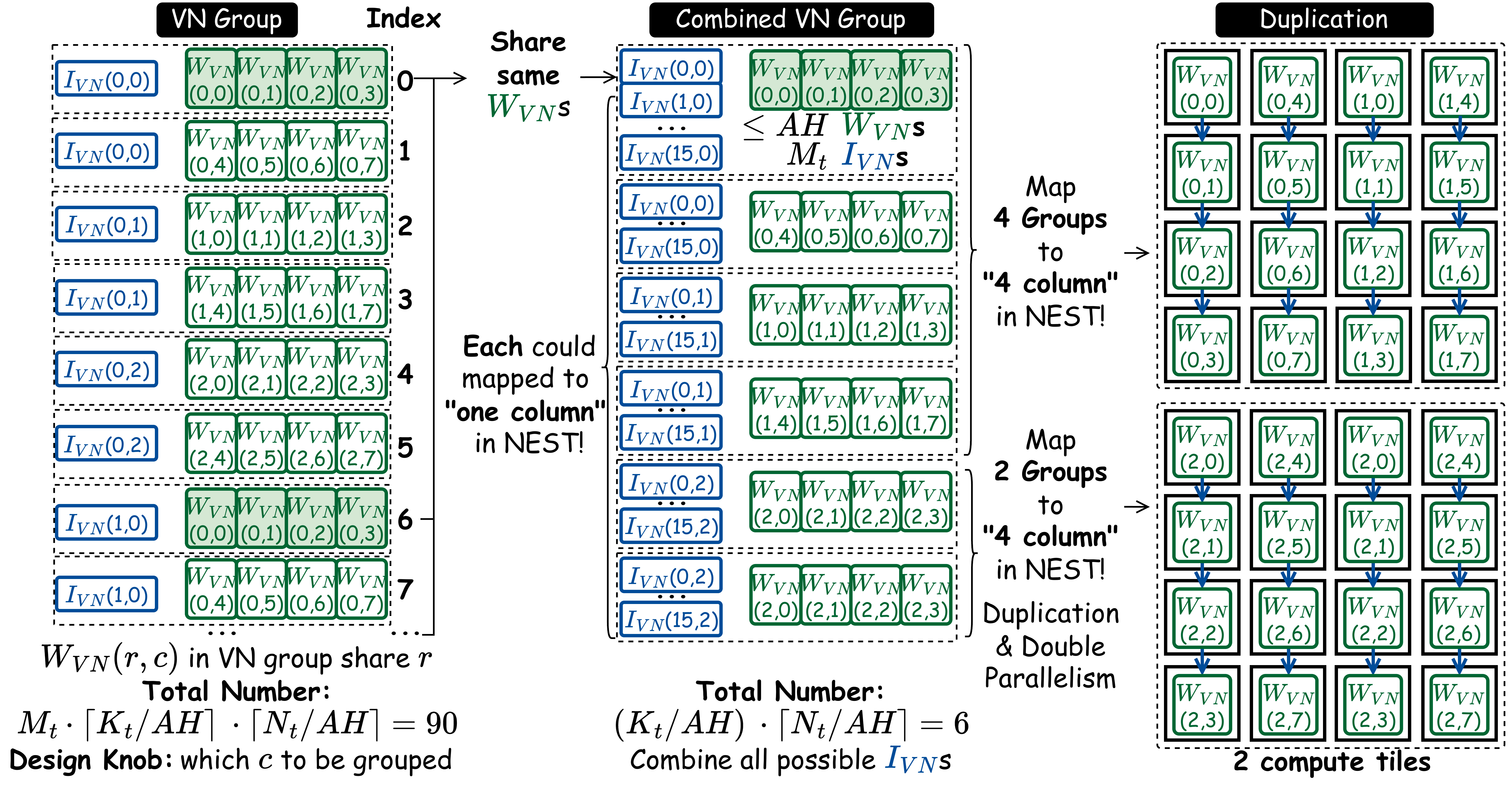}
  \vspace{-6mm}
  \caption{Mapper illustration of ``VN Group", ``Combination" and ``Duplication" to lower \ding{183} in \figref{fig:workload_case_study} to two compute tiles.}
  \label{fig:mapper}
  \vspace{-3mm}
\end{figure}

The FEATHER+ mapper performs \emph{mapping-first, layout-second} search. It first enumerates candidate mappings, and then derives buffer layouts that realize each mapping without bank conflicts. This ordering is motivated by a key observation: although the full joint space of mapping and layout is large, the mapping space itself can be compactly parameterized by only three knobs: (1) compute-tile size, (2) VN-group formation, and (3) column duplication. Once a mapping candidate is fixed, the remaining layout search is restricted to layout \emph{orders} and level-0 \emph{partition factors}, which is much smaller and can be checked efficiently for feasibility. Among all feasible \{mapping, layout\} pairs, the mapper returns the one with minimum estimated latency. \figref{fig:workload_case_study} and \figref{fig:mapper} illustrate this flow using a matrix multiplication example lowered into two compute tiles. From the mapper's perspective, IO-S is equivalent to a transposed WO-S configuration.

\subsubsection{Step 1 - Lower the Workload into Virtual Neurons}

The input workload, such as convolution or matrix multiplication, is first rewritten into a set of \emph{Virtual Neurons} (VNs), e.g., \ding{183} in \figref{fig:workload_case_study}. Each tensor is partitioned along its reduction dimension into fixed-length VNs, where each VN has length at most $AH$ so that it fits the per-PE local registers. VN is the canonical software abstraction manipulated by the mapper.

Under this representation, each tensor becomes a 2D VN array indexed by one non-reduction rank and one reduction-tile rank, i.e., 2D logical arrays of \ivn{}, \wvn{}, and \ovn{}. Execution then reduces to matching one \ivn{} and one \wvn{} that share the same reduction-rank index onto the same PE for a dot product. For instance, \ivn{}($\ast,0$) can only pair with \wvn{}($0,\ast$), where $\ast$ denotes any non-reduction index. In other word, workloads become a logical GEMM ($
{\color{myred}{O}}_{m,n}^{P=M,Q=N}
=
{\color{myblue}{I}}_{m,k}^{M=M,J=K}
\cdot
{\color{mygreen}{W}}_{k,n}^{K=K,N=N}
$) with rank shape $(M, K, N)$.

\subsubsection{Step 2 - Tile the Workload}

Tiling serves two purposes: it bounds tensors within the on-chip buffer capacity and exposes control to tuning the tile sizes for overlapping off-chip transfers with computation. For GEMM, a tile therefore represents a subproblem of shape $M_t \times K_t \times N_t$ to be executed by one or more NEST invocations.

\subsubsection{Step 3 - Form VN Groups}

Each NEST column is an independent mapping unit consisting of $AH$ PEs, with each PE storing one \wvn{} under WO-S or \ivn{} under IO-S. The mapper therefore groups VNs into \emph{VN groups}, where each VN group contains one streaming \ivn{} and up to $AH$ \wvn{}s that can consume that same \ivn{} when mapped onto one column. A VN group is thus the smallest schedulable unit for a single column. For a GEMM tile, the number of VN groups is
$
M_t \cdot \Big\lceil \frac{K_t}{AH} \Big\rceil \cdot \Big\lceil \frac{N_t}{AH} \Big\rceil.
$
Within a VN group, all \wvn{}s share the same reduction-rank index, but the mapper may choose different non-reduction indices to co-locate in the same group. This choice defines the second mapping knob.

\subsubsection{Step 4 - Combine VN Groups Across Streaming Inputs}

VN groups are then merged to capture reuse of stationary \wvn{}s across multiple streamed \ivn{}s. A \emph{combined VN group} contains all \ivn{}s that reuse the same set of up to $AH$ \wvn{}s when executed on one NEST column. Equivalently, it specifies the full workload assigned to one column before moving to a different stationary-\wvn{} configuration.

For a GEMM tile, the number of combined VN groups is $\Big\lceil \frac{K_t}{AH} \Big\rceil \cdot \Big\lceil \frac{N_t}{AH} \Big\rceil$. 
All \ivn{}s and \wvn{}s in the same combined VN group share the same reduction-tile index. This step makes explicit the reuse opportunity that would otherwise be hidden across many individual VN groups.

\subsubsection{Step 5 - Select Column Duplication}

Since NEST has $AW$ columns, the mapper must decide how many combined VN groups are executed concurrently and whether some of them should be duplicated across columns. At one extreme, the mapper selects $AW$ distinct combined VN groups and assigns one to each column. At the other extreme, it may select fewer than $AW$ groups and replicate their \wvn{}s across multiple columns, allowing different subsets of \ivn{}s from the same group to be processed in parallel. This replication factor is the third mapping knob.

For example, if a tile produces six combined VN groups and $AW=4$, the mapper may schedule four distinct groups in the first invocation, then schedule the remaining two groups with duplication factor two in the second invocation. This choice improves column utilization when the number of available groups does not evenly match the physical column count, and it highlights why FEATHER+ benefits from reconfigurable mappings even within a single logical tile.

\subsubsection{Step 6 - Search for Feasible Layouts}

The three mapping knobs above generate a pool of candidate mappings. For each mapping candidate, the mapper searches for tensor layouts that realize the candidate without bank or port conflicts. The layout search is restricted to layout orders and level-0 partition factors, and each candidate is checked against the following feasibility conditions.

\paragraph{Buffer-capacity legality}
The \ivn{}, \wvn{}, and \ovn{} must fit within/com streaming, stationary, and output buffers.

\paragraph{Streaming/stationary-buffer legality}
A combination of \texttt{ExecuteMapping}, \texttt{SetIVNLayout}, \texttt{SetWVNLayout} must not create bank conflicts when reading \ivn{}s or \wvn{}s.

\paragraph{Output-buffer legality}
The chosen \texttt{SetOVNLayout} must not introduce output-buffer port conflicts under.

Any candidate that violates above conditions is discarded.

\subsubsection{Step 7 - Generate the MINISA ISA Trace}

Among all feasible \{mapping, layout\} pairs, the mapper selects the one with minimum estimated end-to-end latency. The selected solution is then deterministically lowered into a MINISA ISA trace consisting of \texttt{ExecuteMapping}, \texttt{SetWVNLayout}, \texttt{SetIVNLayout}, and \texttt{SetOVNLayout} instructions. The resulting trace is finally evaluated by the FEATHER+ analytical performance model to estimate latency and utilization.

For multi-layer workloads, the mapper additionally enforces inter-layer layout compatibility: the output layout of layer $i$ must match the input layout expected by layer $i+1$. It then searches over all surviving cross-layer combinations and selects the choice with minimum overall latency.
\section{Evaluation}
\label{sec:evaluation}

% Our evaluation addresses four questions:
% \textbf{(Q1)} How much control overhead does MINISA eliminate?
% \textbf{(Q2)} How much end-to-end performance does MINISA unlock for FEATHER+ across AI, HE, and ZKP workloads?
% \textbf{(Q3)} What's the impact of $AH$/$AW$ in performance and utilization?
% \textbf{(Q4)} What hardware overhead does FEATHER+ introduce relative to FEATHER?

\subsection{Evaluation Setup}
\label{sec:eval_setup}

% \begin{table}[t]
% \centering
% \scriptsize
% \caption{Experimental setup for FEATHER+.%Data SRAM is split 40\%/40\%/20\% across streaming (StrB)/stationary (StaB)/output buffers (OB). Instruction buffer is dedicated. 
% % Bandwidths are in \emph{bytes/cycle} (equivalently GB/s at 1\,GHz under our model).
% }
% \label{tab:minisa_setup}
% \resizebox{\columnwidth}{!}{
% \begin{tabular}{cc|ccc|c|cccc}
% \hline
% \multirow{2}{*}{$AH$} & \multirow{2}{*}{$AW$} & 
% \multicolumn{3}{c|}{On-chip capacity (MB)} &
% \multirow{2}{*}{\makecell{Peak \\ MAC/cycle}} &
% \multicolumn{4}{c}{Bitwidth} & 
% \cline{3-5}\cline{7-10}
%  & & StrB/ StaB & OB & Instr. &  & {\color{myblue}$I$} & {\color{mygreen}$W$}& {\color{myred}$O$} & Instr. \\
% \hline
% 4 & 4/16/64 & 1.6 & 0.8 & 0.5 & 16 & 4 & 4 & 16  & 9 \\
% % 4 & 16 & 1.6 & 0.8 & 0.5 & 64 & 8 & 8 & 32  & 9 \\
% % 4 & 64 & 1.6 & 0.8 & 0.5 & 256 & 16 & 16 & 64 & 9 \\
% 8 & 8/32/128 & 6.4 & 3.2 & 1.0 & 64 & 32 & 32 & 128 & 9 \\
% % 8 & 32 & 6.4 & 3.2 & 1.0 & 256 & 64 & 64 & 256 & 9 \\
% % 8 & 128 & 6.4 & 3.2 & 1.0 & 1024 & 64 & 64 & 256 & 9 \\
% 16 & 16/64/256 & 25.6 & 12.8 & 2.0 & 256 & 128 & 128 & 512 & 9 \\
% % 16 & 64 & 25.6 & 12.8 & 2.0 & 1024 & 128 & 128 & 512 & 9 \\
% % 16 & 256 & 25.6 & 12.8 & 2.0 & 4096 & 128 & 128 & 512 & 9 \\
% \hline
% \end{tabular}}
% \end{table}

\textbf{Methodology.} We use a two-level simulation framework. A cycle-accurate RTL model of FEATHER+ (supporting both MINISA and micro-instructions) validates functional correctness and timing. The RTL-verified FEATHER+ mapper with an analytical model extends the study to larger array scales.

\textbf{Configuration.} We use FEATHER+ with $(AH, AW)$ sweeping $\in \{(4,4/16/64), (8, 8/32/128), (16, 16/64/256)\}$. On-chip data SRAM scales with $AH$ and is partitioned into streaming (40\%), stationary (40\%), and output (20\%) buffers. A dedicated instruction buffer (0.5\,MB, 1\,MB, 2\,MB) is served by a fixed off-chip instruction interface of 9\,B/cycle. Off-chip data bandwidths are modeled as $AW$ B/cycle for inputs/weights and $4AW$ B/cycle for outputs.

\textbf{Workloads.} We evaluate 50 GEMM kernels from three domains: LLM inference (GPT-OSS 20B), Homomorphic Encryption (FHE) bootstrapping\cite{openfhe}, and Number-Theoretic Transform (NTT) kernels from both FHE\cite{tong2025CROSS} and Zero-Knowledge Proof (ZKP)\cite{tong2025MORPH}. Shapes are listed in \tabref{tab:minisa_workloads}. \label{sec:workload}

\begin{table}[t]
\centering
\scriptsize
\caption{Workload specifications of matrix multiplication.}
\vspace{-2mm}
\label{tab:minisa_workloads}
\resizebox{\columnwidth}{!}{
\begin{tabular}{l|l}
\hline
Workload & Matrix Multiplication of Shape $O_{m,n}^{P=M,Q=N} = I_{m,k}^{M=M,J=K} \cdot W_{k,n}^{K=K,N=N}$ \\
\hline
FHE: BConv & $(65536{\times}K)\cdot(K{\times}N)$, $K\in[28,60]$, $N\in[72,160]$ (41 shapes) \\
FHE: NTT & $J=K=N\in\{1024,2048,4096\}, M\in\{64,128,256\}, M\le K/16$ \\
ZKP: NTT & $J=K=N\in\{8192,16384,32768\}, M\in\{K/32,K/16\}$ \\
GPT-oss & $M=2048,(J=K,N)\in\{(64,2048),(2880,4096/5120/201088),(4096,2880)\}$ \\ 
\hline
\end{tabular}}
\vspace{-2mm}
\end{table}

\begin{table}[t]
\centering
\scriptsize
\caption{Experimental setup for FEATHER+.}
\vspace{-2mm}
\label{tab:minisa_setup}
\resizebox{\columnwidth}{!}{
\begin{tabular}{c|c|ccc}
\hline
\multirow{2}{*}{($AH$, $AW$)} & 
On-chip capacity (MB) &
\multicolumn{3}{c}{MINISA ISA Bitwidth} \\
\cline{2-2}\cline{3-5}
 & (StrB/ StaB, OB, Instr.) & \texttt{Set*VNLayout} & \texttt{E.Mapping} & \texttt{E.Streaming} \\
\hline
(4, 4/16/64) & (1.6, 0.8, 0.5) & 42/40/38 bits & 81/83/85 bits & 57/51/45 bits \\
(8, 8/32/128) & (6.4, 3.2, 1.0) & 43/41/39 bits & 86/88/90 bits & 58/52/46 bits \\
(16, 16/64/256) & (25.6, 12.8, 2.0) & 44/42/40 bits & 91/93/95 bits & 59/53/47 bits \\
\hline
\end{tabular}}
\end{table}

\textbf{Baselines.} For each workload and array size, we compare the following two configurations under identical mappings.

\noindent $\bullet$ \textbf{Micro-instruction (Baseline):} Explicit, fine-grained control where every switch and PE is configured per cycle.

\noindent $\bullet$  \textbf{MINISA (Ours):} VN-based MINISA instructions.

\textbf{Metrics.} (1) \textit{Latency}: GPU/TPU latencies are measured with Nsight/JAX-profiler on realistic traces; FEATHER+ results come from cycle-accurate RTL. We report best-latency from: CUDA kernels with tiled/strided/continguous layouts on GPU; and best sharding of (M, N) over eight tensor-cores(TPU). (2) \textit{Utilization}: average compute utilization of the entire end-to-end execution of matrix multiplications.

\subsection{Control Overhead Evaluation}
\label{sec:eval_control}
\begin{figure*}[!t]
    \centering
    % \vspace{-3mm}
    \includegraphics[width=\textwidth]{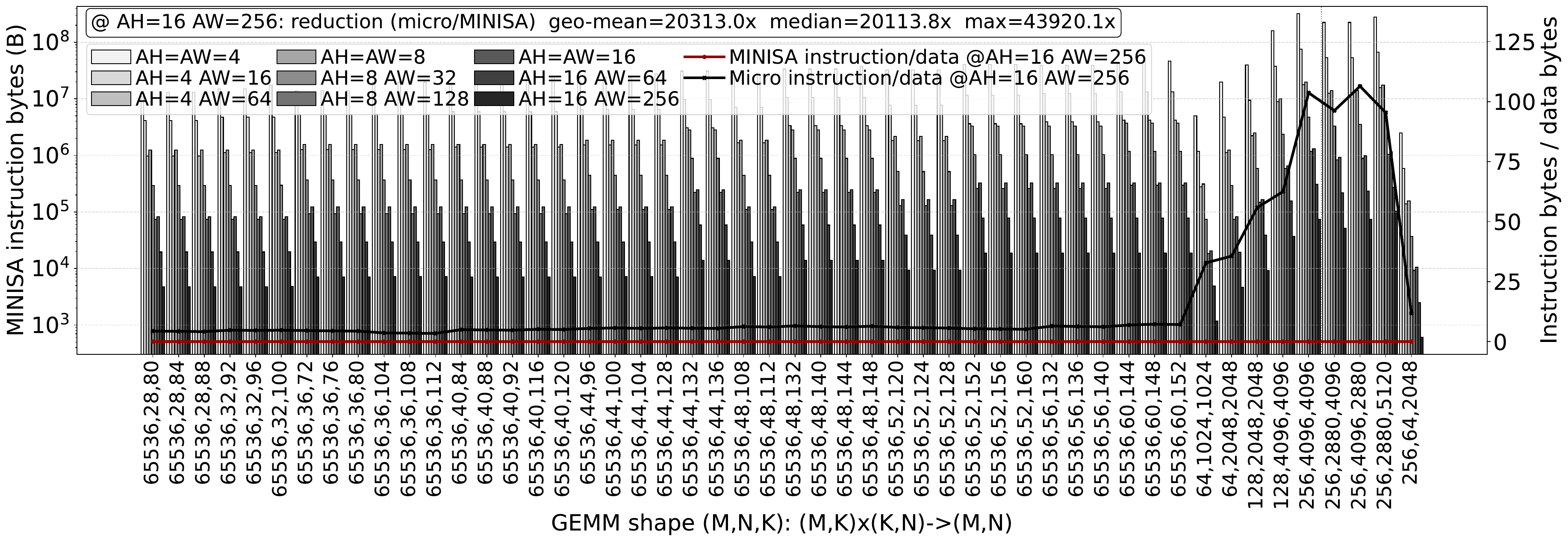}
    \vspace{-7mm}
    \caption{Instruction-overhead reduction. At $16{\times}256$, MINISA lowers instruction bytes by a geometric mean of $(2{\times}10^5){\times}$ versus micro-instructions (bar chart). Black and red lines show instruction-to-data ratio for baseline and MINISA, respectively. \textbf{Takeaway:} MINISA shrinks instruction volume by at most $4.4{\times}10^5{\times}$, removing instruction loading from the critical path.}
    \label{fig:instr_reduction}
    \vspace{-4mm}
\end{figure*}

We first quantifies and compares total instruction bytes and fetching latency for MINISA against the baseline in \figref{fig:instr_reduction}.

\subsubsection{Instruction Storage Reduction}

Across all 50 workloads, micro-instructions could take up-to 100$\times$ storage overhead than data. Because the micro-instruction stream must explicitly configure switch and addresses fabric. MINISA reduces such costs by a geometric mean of \textbf{$2{\times}10^4{\times}$} at $(AH,AW)=(16,256)$. The reduction grows roughly with workload sizes as more instructions are needed for bigger workloads. 

% \textbf{Instruction-to-Data Ratio.} The ratio of instruction traffic to data traffic shows the bottleneck directly. At $(AH,AW)=(16,256)$, the micro-instruction baseline fetches \textbf{$49{\times}$} more instruction than data (geometric mean), so control traffic dominates off-chip bandwidth. Under MINISA, instruction traffic becomes negligible at only \textbf{$2.6{\times}10^{-4}$} of data volume.

\subsubsection{Instruction Fetch Latency Reduction}
\begin{figure}[!t]
    \centering
    % \vspace{-3mm}
    \includegraphics[width=\columnwidth]{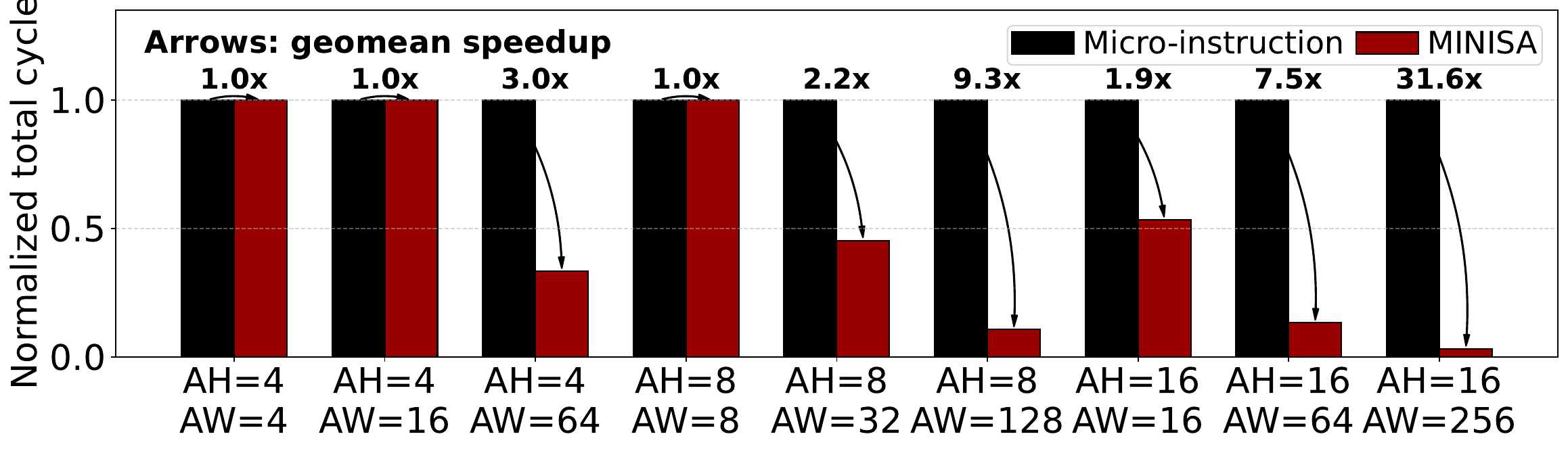}
    \vspace{-6mm}
    \caption{End-to-end speedup and stall analysis. MINISA achieves up to $31.6{\times}$ speedup as array size scales to $16{\times}256$. \textbf{Takeaway:} The performance gain stems from eliminating instruction-fetch stalls. The micro-instruction baseline becomes instruction-bound at large scales, which MINISA reduces to zero instruction-fetch stalls across all configurations.}
    \label{fig:speedup_over_micro_instr}
    \vspace{-2mm}
\end{figure}

We evaluate how instruction compression translates into end-to-end speedup (\figref{fig:speedup_over_micro_instr}) for FEATHER+ with different scales.

\squishlist
\item \textbf{Small scale ($\leq\!64$ PEs):} At $(AH,AW)\!=\!(4,4),\!(4,16),$ $\!(8,8)$, micro-instructions consumption is small enough to be hidden by compute, making speedup as 1.
\item \textbf{Large scale ($>\!64$ PEs):} As arrays grow, BIRRD instructions and buffer addresses grow by $O(AW\log_2(AW))$ and $O(D\!\times\!AW)$. Stall cycles from instruction fetch rise from \textbf{75.3\%} at $(4,64)$ to \textbf{96.9\%} at $(16,256)$ in baseline.
\squishend

Because MINISA removes fetch stalls across all tested sizes, geometric-mean speedup increases with scale: \textbf{$1.9{\times}$} at $16{\times}16$, \textbf{$7.5{\times}$} at $16{\times}64$, and up to \textbf{$31.6{\times}$} at $16{\times}256$. These results proves the efficacy of VN abstraction in concealing rich reconfigurations in hardware with succinct ISAs.
\subsection{Performance Evaluation}
\label{sec:eval_perf}

\subsubsection{Comparison against industry baselines (TPU\&GPU)}

\begin{figure}[!t]
    \centering
    % \vspace{-3mm}
    \includegraphics[width=\columnwidth]{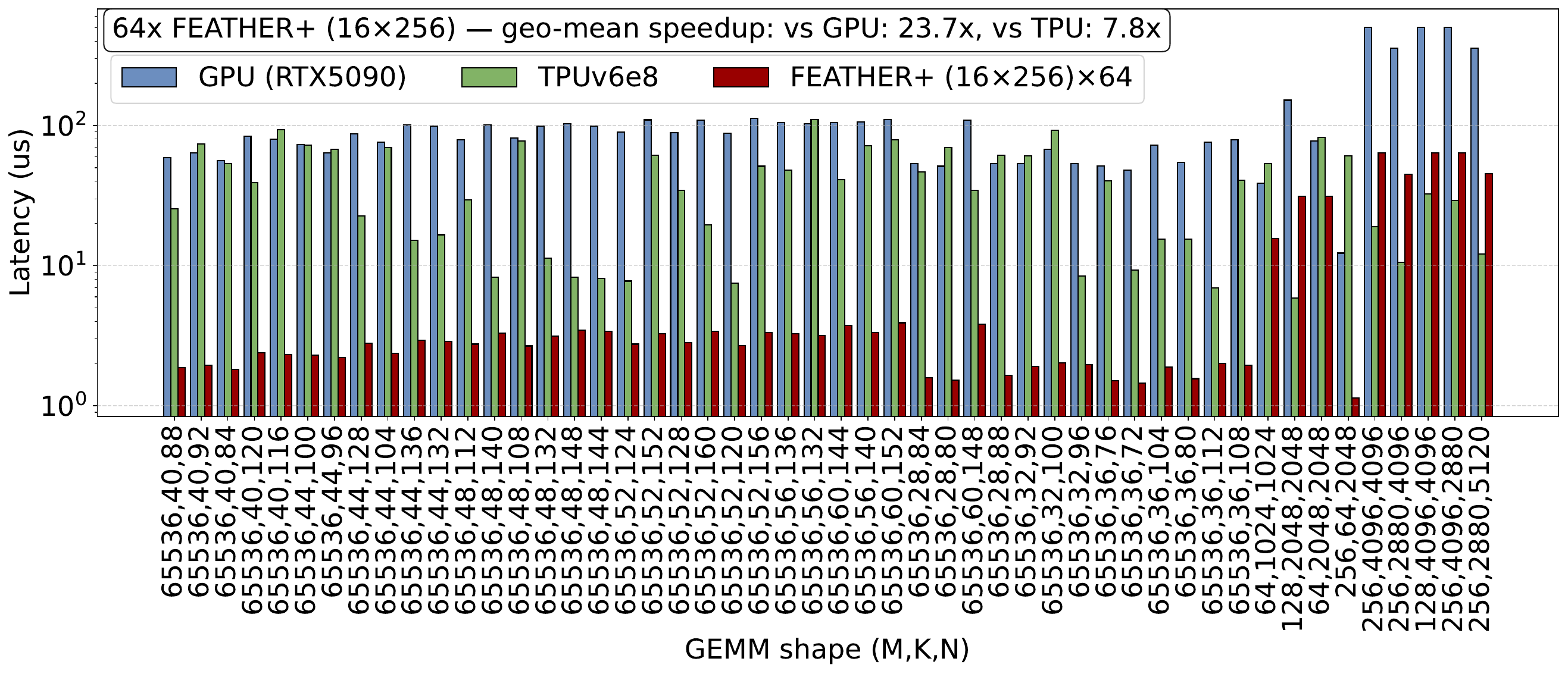}
    \vspace{-6mm}
    \caption{Latency comparison among GPU (RTX 5090), TPUv6e-8 (256$\times$256$\times$8), and FEATHER+ ((16$\times$256)$\times$64).}
    \label{fig:latency_vs_gpu_tpu}
    \vspace{-2mm}
\end{figure}

We compare FEATHER+ vs. RTX5090 and TPUv6e. All devices are scaled to roughly the same power budget of 575W: eight TPUv6e tensor cores and 64 instances of FEATHER+ 16$\times$256 connected as a 8$\times$8 mesh, as shown in \figref{fig:latency_vs_gpu_tpu}.

FEATHER+ achieves 23.7$\times$ (vs. RTX5090) and 7.8$\times$ (vs. TPUv6e) geo-mean speedup. The key reason is granularity mismatch in less-flexible GPUs/TPUs (e.g., for INT8, TPUv6e process GEMM at minimal granularity of 8$\times$256$\times$256, while RTX5090 does it at 16$\times$32$\times$8). When GEMM shape does not divide these fixed granularity, compute will be underutilized. FEATHER+ supports matrix multiplication at the granularity of $T\times AH\times AH$ per PE column to align diverse shape of GEMM in runtime, where $T\in[1, M]$ is total number of \ivn{}s to be streamed per row.

\begin{figure}[!t]
    \centering
    % \vspace{-3mm}
    \includegraphics[width=\columnwidth]{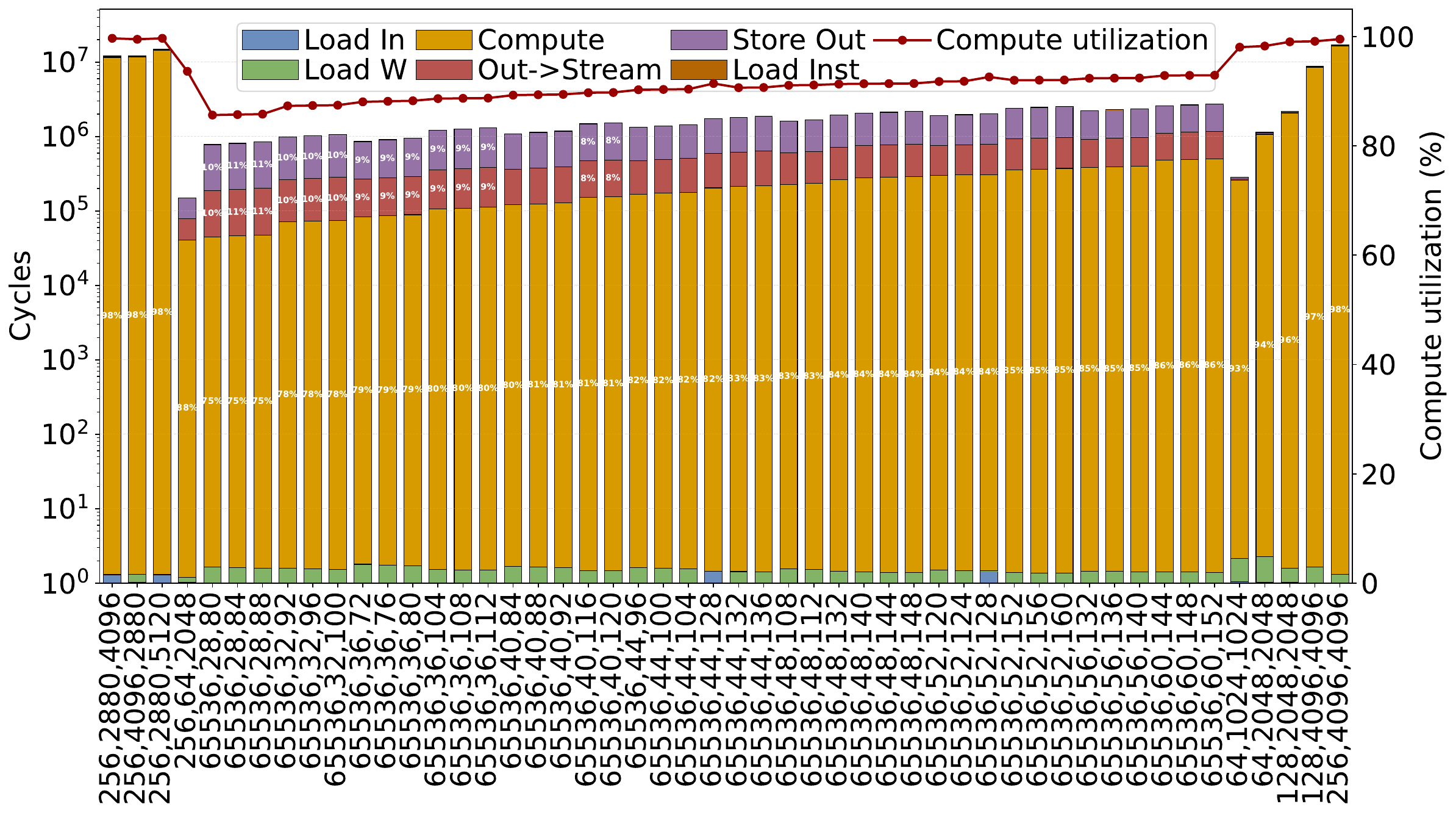}
    \includegraphics[width=\columnwidth]{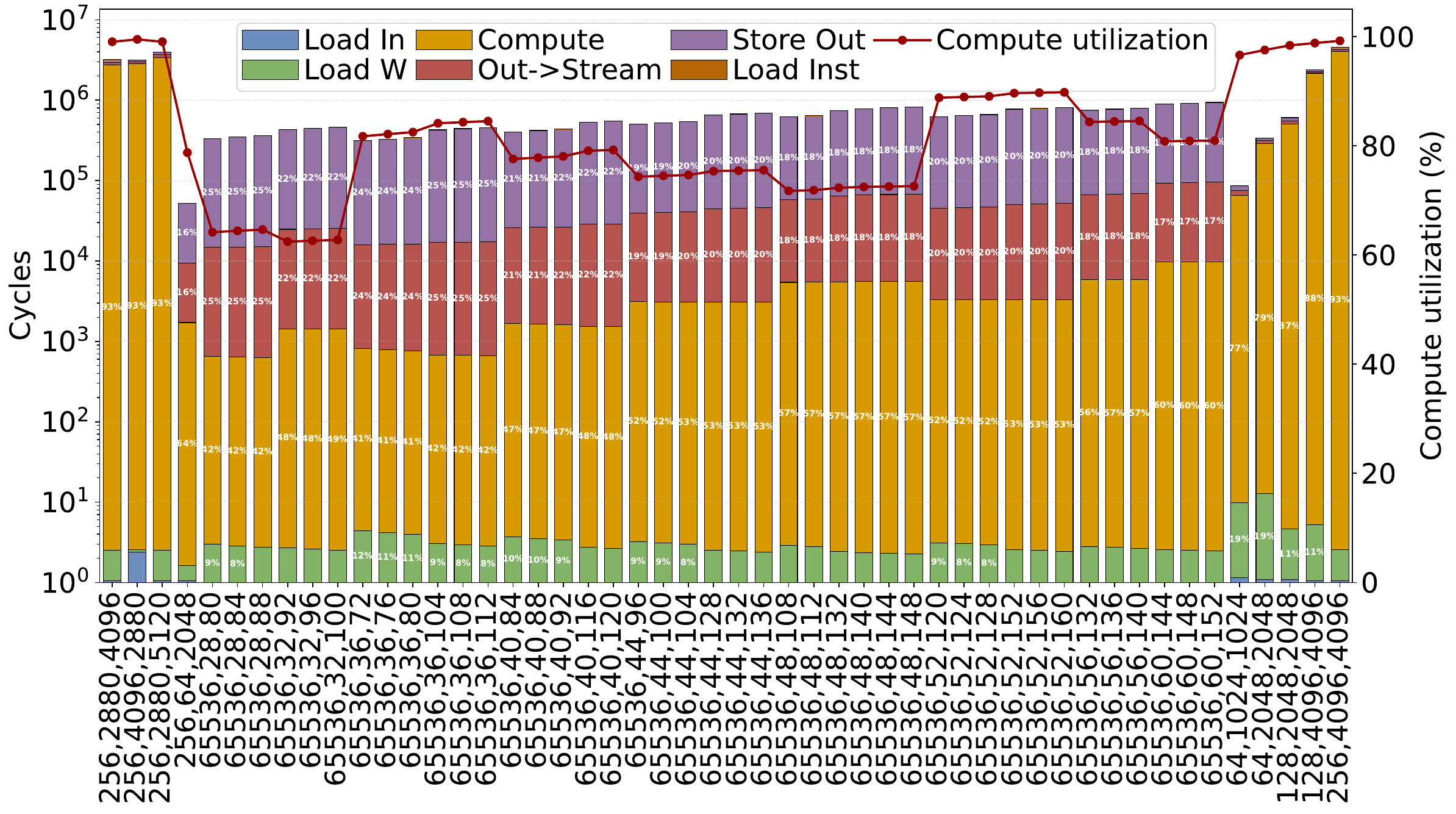}
    \includegraphics[width=\columnwidth]{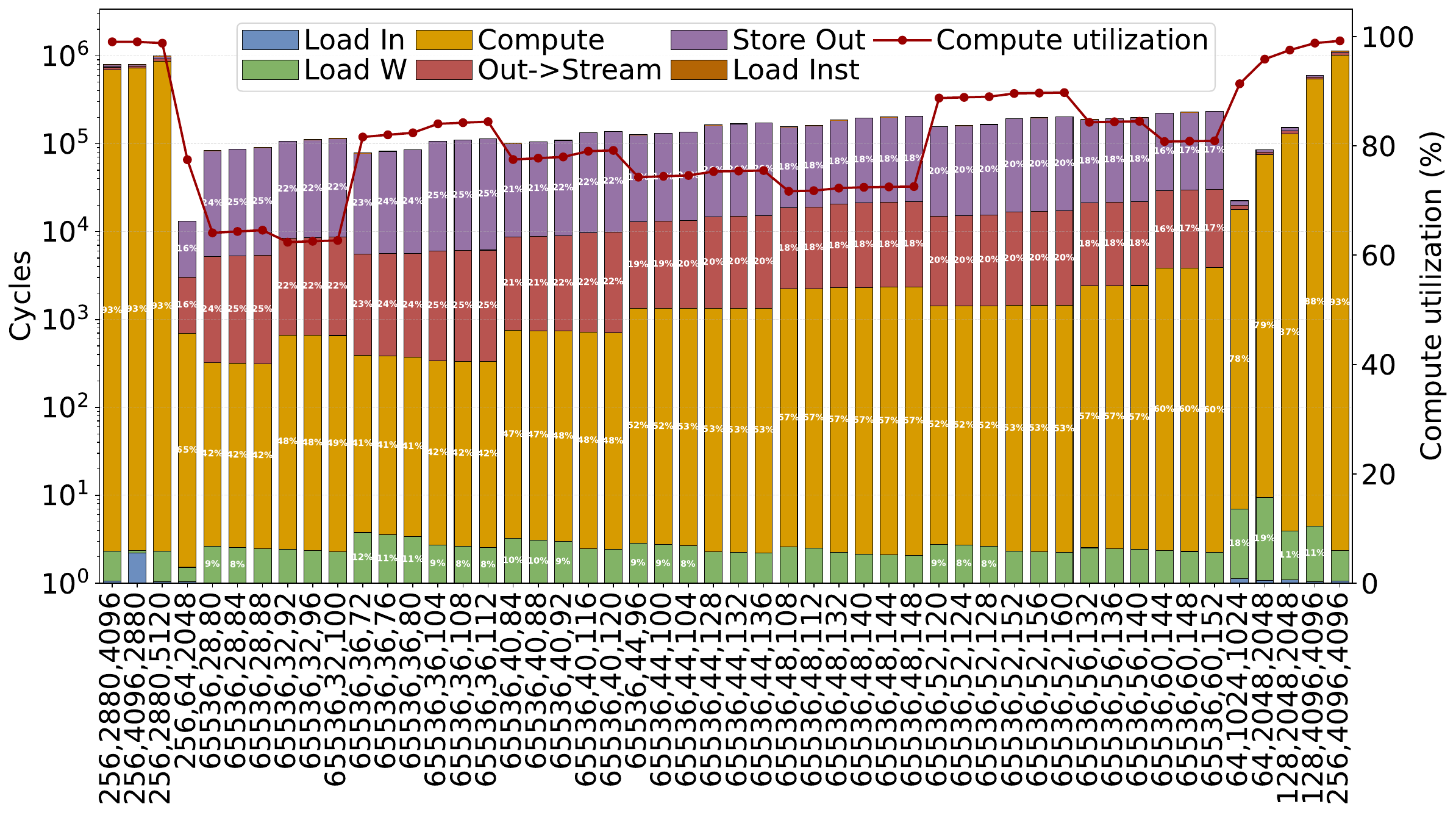}
    \vspace{-6mm}
    \caption{Latency breakdown and compute utilization of representative workloads. \textbf{Takeaway}: FEATHER+ (4$\times$64, upper chart), (16$\times$64, middle chart) and (16$\times$256, lower chart) with MINISA could successfully utilize available PEs for all irregular workload shapes, such as $K{=}10$ and $K={2^n}, n\in \mathbb{Z}^+$.}
    \label{fig:latency_compute_utilization_nest16}
    % \vspace{-2mm}
\end{figure}

\subsubsection{Latency Breakdown}
We further dissect architectural efficiency using cycle-level breakdowns of representative workloads in \figref{fig:latency_compute_utilization_nest16}. We separate execution time into four overlapping components: \emph{Compute}, input/weight streaming (\emph{Load In/W}), output movement from the output buffer to the Streaming/Stationary Buffer (\emph{Out$\rightarrow$Stream}), and off-chip output transfer (\emph{Store Out}).

\textbf{Robustness to Irregular Shapes.} The compute-utilization curve (red line in \figref{fig:latency_vs_gpu_tpu}) remains consistently high across a wide range of tensor shapes, indicating that FEATHER+ is largely robust to shape irregularity.

\noindent $\bullet$ \textit{Irregular workloads:} For FHE and ZKP kernels with shape not dividing the compute granularity (e.g., $K{=}10$ or $N{=}21$), FEATHER+ still sustains $>60\%$ average compute utilization. This is in sharp contrast to rigid systolic arrays with 3\% compute utilization, where mismatched dimensions typically introduce padding overhead and leave a majority of PEs idle.

\noindent $\bullet$ \textit{Regular workloads:} For GEMMs whose dimensions align exactly with TPUv6e execution granularity (e.g., $K, N \in \{1024, 2048\}$), all architectures can approach peak utilization. In this regime, FEATHER+ is about 30\% slower due to reconfiguration overhead rather than underutilization. Even so, this overhead remains lower than that of RTX5090, demonstrating that MINISA effectively amortizes reconfiguration through a compact control path and low instruction overhead.

\textbf{Takeaway:} Reconfigurability allows FEATHER+ to preserve high compute utilization even for highly irregular workloads, where rigid architectures suffer from padding-induced inefficiency. MINISA further keeps the cost of this flexibility low by minimizing reconfiguration overhead.

\subsection{Scalability Ablation Study}
We analyze how performance and resources of FEATHER+ scale with array width ($AW$) and array height ($AH$).

\subsubsection{Scaling $AW$: More Independent Parallelism}

$AW$ is the number of NEST columns. Because columns operate independently, increasing $AW$ adds column-level parallelism without changing per-column execution granularity.

With $AH{=}16$, scaling $AW$ from $64$ to $256$ delivers nearly linear scaling: FEATHER+ achieves about $4\times$ average speedup with almost unchanged compute utilization across benchmarks (\figref{fig:latency_compute_utilization_nest16}). Thus, wider arrays translate directly into throughput rather than additional underutilization.

The hardware cost of scaling $AW$ is also favorable. NEST and on-chip buffers grow linearly, i.e., $O(AW)$. Interconnect overhead grows faster but remains subquadratic: BIRRD scales as $O(AW \log AW)$, while the distribution network is bounded by $O(AW^2)$. Overall, scaling $AW$ is an efficient way to increase throughput, with cost dominated by interconnect.

\subsubsection{Scaling $AH$: Higher Parallelism with Temporal Reuse Restriction}
Increasing $AH$ grows overall MACs and parallelism but also increases workload granularity for fully utilizing one PE column to $1\!\times\!AH\!\times\!AH$ matrix multiplication.

Specifically, each PE performs an $AH$-element local dot product. $AH$ determines the maximum VN size supported by each PE, and thus sets FEATHER+'s compute granularity. All PE in a column reuses the same streaming tensor, therefore scaling $AH$ adding more temporal reuse restriction.

Full utilization of an $AH\!\times\!AW$ array requires $VN_{\text{size}}{=}AH$. When $VN_{\text{size}}<AH$, FEATHER+ activates only $VN_{\text{size}} \times AW$ PEs and uses only $VN_{\text{size}}$ local registers per PE, skipping the remaining rows to reduce stationary tensor loading latency and pipeline latency.

This tradeoff appears in performance. With $AW{=}64$, scaling $AH$ from $4$ to $16$ yields $2.6\times$--$4\times$ speedup, depending on workload size. The gain comes from larger dot products and higher intra-column parallelism, but a larger $AH$ also raises the minimum workload granularity needed for full utilization. Hence, scaling $AH$ improves peak throughput but makes utilization more sensitive to VN size.

Resource-wise, local storage grows quadratically, i.e., $O(AH^2)$; number of multipliers scale linearly, i.e., $O(AH)$.

\textbf{Takeaway.} Increasing $AW$ scales independent parallelism. Increasing $AH$ raises parallelism with temporal reuse requirement and increases \emph{compute granularity}. %In short, $AW$ scales \emph{parallelism}, while $AH$ scales both \emph{parallelism} and \emph{granularity}.

\subsection{Resource Overhead Evaluation}
\label{sec:eval_resource}

Finally, we quantify the hardware cost of FEATHER+ architectural refinements (including all-to-all distribution) needed to support dynamic workloads. \tabref{tab:resource_cmp} summarizes area results.

Compared to FEATHER\cite{tong2024FEATHER}, FEATHER+ only adds up-to 7\% resources overhead because the area overhead of introduced all-to-all distribution network are amortized over distributed register and compute resources.

\begin{table}[t]
\centering
\scriptsize
% \vspace{-2mm}
\setlength{\tabcolsep}{3.5pt}
\caption{Post-PnR area ($\mu m^2$) and power (mW) comparison FEATHER (F) vs. FEATHER+ (F+) in TSMC 28nm.}
\vspace{-3mm}
\label{tab:resource_cmp}
\begin{tabular}{ccccccc}
\hline
\multirow{2}{*}{\textbf{Setup}} &
\multicolumn{3}{c}{\textbf{Area} ($\mu m^2$)} &
\multicolumn{3}{c}{\textbf{Power} (mW)} \\
\cline{2-4} \cline{5-7}
& \textbf{F} & \textbf{F+} & \textbf{Increase $\uparrow$} & \textbf{F} & \textbf{F+} & \textbf{Increase  $\uparrow$} \\
\hline
$4{\times}4$     & 70598   & 71573   & 1.38\% & 44.59   & 45.34   & 1.67\% \\
$8{\times}8$     & 174370  & 176573  & 1.26\% & 108.97  & 110.49  & 1.39\% \\
$16{\times}16$   & 476174  & 482044  & 1.23\% & 293.47  & 297.72  & 1.45\% \\
$4{\times}64$    & 1259903 & 1352697 & 7.37\% & 854.77  & 915.14  & 7.06\% \\
$8{\times}128$   & 3198595 & 3441146 & 7.58\% & 2240.27 & 2350.88 & 4.94\% \\
% $16{\times}256$   & ?? & 10239004 & ??\% & ?? & 6983 & ?\% \\
\hline
\end{tabular}
\\ We fix the depth of all on-chip buffers to 64 and implement them as registers for PnR. In a practical deployment, larger buffers should instead use SRAM macros.
% \vspace{-5.5mm}
\end{table}

% In summary, MINISA removes the control-scaling bottleneck in reconfigurable accelerators. By moving from wire-level micro-control to VN-level abstraction, it cuts control traffic by up to five orders of magnitude and delivers nearly $100{\times}$ end-to-end speedup at large scales, while adding less than 1\% area overhead for practical array sizes.

\section{Related Work}
\label{sec:relatedWork}

\textbf{Instruction Set for AI Accelerators.} 
Commercial AI accelerators typically employ monolithic instruction sets driving fixed functional units to maximize efficiency for standard deep learning operators. Architectures like Google's TPU~\cite{tpuv4i}, Meta's MTIA~\cite{MTIA}, and Amazon's Trainium~\cite{nki_isa} are designed for coarser-grained regular matrix multiplication. These rigid gran lack the fine-grained distinct programmability required to support the irregular dataflows and complex algebraic structures found in emerging workloads like Homomorphic Encryption (HE) and Zero-Knowledge Proofs (ZKP).

\textbf{Reconfigurable Dataflow Accelerators.} 
To address the diversity of tensor shapes and operators, reconfigurable architectures such as MAERI~\cite{kwon2018maeri}, SIGMA~\cite{sigma_eq}, and Flexagon~\cite{flexagon} introduce flexible interconnects to support diverse dataflows~\cite{parashar2019timeloop}. Further efforts like PolyGraph~\cite{polygraph}, DSA-Gen~\cite{DSAGen}, and Over-Gen~\cite{over_gen} explore overlay generation to map varied kernels. However, these designs often decouple computation from memory layout, incurring significant data marshalling overheads. FEATHER~\cite{tong2024FEATHER} addresses this by enabling low-cost co-switching of dataflow and data layout. 
\textit{Crucially, however, prior reconfigurable baselines—including FEATHER—rely on fine-grained micro-configuration.} As array sizes scale, this results in excessive control overhead and instruction-fetch stalls that bottleneck end-to-end performance. MINISA specifically targets this scalability crisis by raising the abstraction level to Virtual Neurons, eliminating redundant control traffic while preserving architectural flexibility.

\textbf{AI ASICs for Cryptography (HE \& ZKP).} 
The high computational cost of cryptographic primitives has driven a bifurcation in hardware support. Prior works have demonstrated the superiority of leveraging AI ASICs for HE\cite{tong2025CROSS} and ZKP primitives\cite{tong2025MORPH} by converting high-precision modular arithmetic down to low-precision dense matrix multiplication with diverse shapes. 
MINISA and FEATHER+ improves the compute utilization for small matrix multiplications which are not aligned with the dimensions of hardware computation.

\section{Conclusion}
\label{sec:conclusion}
MINISA is the first ISA that enables runtime co-switching of dataflow and data layout with negligible instruction overhead, allowing one accelerator to efficiently support workloads with widely varying and irregular shapes. Its key insight is the Virtual Neuron (VN) abstraction, which lifts reconfiguration from individual elements to the granularity of the hardware’s atomic dot-product unit. This abstraction exactly captures the reconfigurability of hardware without adding extra redundant costs, and unifies mapping and layout control in a compiler-friendly form. Coupled with the FEATHER+ architecture, MINISA makes reconfiguration tractable in practice and enables efficient matrix multiplication across diverse shapes from various domains such as FHE, ZKP and AI.
\section{Acknowledge}
This work was supported in part by ACE, one of the seven centers in JUMP 2.0, a Semiconductor Research Corporation (SRC) program sponsored by DARPA. We thank Niansong Zhang, Zhiru Zhang, and reviewers for insightful feedbacks.

%%%%%%%%% -- BIB STYLE AND FILE -- %%%%%%%%
\bibliographystyle{IEEEtranS}
\bibliography{refs}

% Generated by IEEEtranS.bst, version: 1.13 (2008/09/30)
\begin{thebibliography}{10}
\providecommand{\url}[1]{#1}
\csname url@samestyle\endcsname
\providecommand{\newblock}{\relax}
\providecommand{\bibinfo}[2]{#2}
\providecommand{\BIBentrySTDinterwordspacing}{\spaceskip=0pt\relax}
\providecommand{\BIBentryALTinterwordstretchfactor}{4}
\providecommand{\BIBentryALTinterwordspacing}{\spaceskip=\fontdimen2\font plus
\BIBentryALTinterwordstretchfactor\fontdimen3\font minus \fontdimen4\font\relax}
\providecommand{\BIBforeignlanguage}[2]{{%
\expandafter\ifx\csname l@#1\endcsname\relax
\typeout{** WARNING: IEEEtranS.bst: No hyphenation pattern has been}%
\typeout{** loaded for the language `#1'. Using the pattern for}%
\typeout{** the default language instead.}%
\else
\language=\csname l@#1\endcsname
\fi
#2}}
\providecommand{\BIBdecl}{\relax}
\BIBdecl

\bibitem{nki_isa}
\BIBentryALTinterwordspacing
{AWS}. (2025) Aws nki instruction set architecture. [Online]. Available: \url{https://awsdocs-neuron.readthedocs-hosted.com/en/latest/nki/api/nki.isa.html}
\BIBentrySTDinterwordspacing

\bibitem{openfhe}
\BIBentryALTinterwordspacing
A.~A. Badawi, J.~Bates, F.~Bergamaschi, D.~B. Cousins, S.~Erabelli, N.~Genise, S.~Halevi, H.~Hunt, A.~Kim, Y.~Lee, Z.~Liu, D.~Micciancio, I.~Quah, Y.~Polyakov, S.~R.V., K.~Rohloff, J.~Saylor, D.~Suponitsky, M.~Triplett, V.~Vaikuntanathan, and V.~Zucca, ``Openfhe: Open-source fully homomorphic encryption library,'' Cryptology ePrint Archive, Paper 2022/915, 2022, \url{https://eprint.iacr.org/2022/915}. [Online]. Available: \url{https://eprint.iacr.org/2022/915}
\BIBentrySTDinterwordspacing

\bibitem{polygraph}
\BIBentryALTinterwordspacing
V.~Dadu, S.~Liu, and T.~Nowatzki, ``Polygraph: exposing the value of flexibility for graph processing accelerators,'' in \emph{Proceedings of the 48th Annual International Symposium on Computer Architecture}, ser. ISCA '21.\hskip 1em plus 0.5em minus 0.4em\relax IEEE Press, 2021, p. 595–608. [Online]. Available: \url{https://doi.org/10.1109/ISCA52012.2021.00053}
\BIBentrySTDinterwordspacing

\bibitem{MTIA}
\BIBentryALTinterwordspacing
A.~Firoozshahian, J.~Coburn, R.~Levenstein, R.~Nattoji, A.~Kamath, O.~Wu, G.~Grewal, H.~Aepala, B.~Jakka, B.~Dreyer, A.~Hutchin, U.~Diril, K.~Nair, E.~K. Aredestani, M.~Schatz, Y.~Hao, R.~Komuravelli, K.~Ho, S.~Abu~Asal, J.~Shajrawi, K.~Quinn, N.~Sreedhara, P.~Kansal, W.~Wei, D.~Jayaraman, L.~Cheng, P.~Chopda, E.~Wang, A.~Bikumandla, A.~Karthik~Sengottuvel, K.~Thottempudi, A.~Narasimha, B.~Dodds, C.~Gao, J.~Zhang, M.~Al-Sanabani, A.~Zehtabioskuie, J.~Fix, H.~Yu, R.~Li, K.~Gondkar, J.~Montgomery, M.~Tsai, S.~Dwarakapuram, S.~Desai, N.~Avidan, P.~Ramani, K.~Narayanan, A.~Mathews, S.~Gopal, M.~Naumov, V.~Rao, K.~Noru, H.~Reddy, P.~Venkatapuram, and A.~Bjorlin, ``Mtia: First generation silicon targeting meta's recommendation systems,'' in \emph{Proceedings of the 50th Annual International Symposium on Computer Architecture}, ser. ISCA '23.\hskip 1em plus 0.5em minus 0.4em\relax New York, NY, USA: Association for Computing Machinery, 2023. [Online]. Available: \url{https://doi.org/10.1145/3579371.3589348}
\BIBentrySTDinterwordspacing

\bibitem{looptree}
M.~Gilbert, Y.~N. Wu, A.~Parashar, V.~Sze, and J.~S. Emer, ``Looptree: Enabling exploration of fused-layer dataflow accelerators,'' in \emph{2023 IEEE International Symposium on Performance Analysis of Systems and Software (ISPASS)}, 2023, pp. 316--318.

\bibitem{jain2025taidl}
\BIBentryALTinterwordspacing
D.~Jain, M.~Frigo, J.~Arora, A.~Pardeshi, Z.~Wang, K.~Patel, and C.~Mendis, \emph{TAIDL: Tensor Accelerator ISA Definition Language with Auto-generation of Scalable Test Oracles}.\hskip 1em plus 0.5em minus 0.4em\relax New York, NY, USA: Association for Computing Machinery, 2025, p. 1316–1333. [Online]. Available: \url{https://doi.org/10.1145/3725843.3756075}
\BIBentrySTDinterwordspacing

\bibitem{act_arxiv}
\BIBentryALTinterwordspacing
D.~Jain, A.~Pardeshi, M.~Frigo, K.~Patel, K.~Khulbe, J.~Arora, and C.~Mendis, ``Act: Automatically generating compiler backends from tensor accelerator isa descriptions,'' Oct. 2025. [Online]. Available: \url{https://arxiv.org/abs/2510.09932}
\BIBentrySTDinterwordspacing

\bibitem{tpuv4i}
N.~P. Jouppi, D.~Hyun~Yoon, M.~Ashcraft, M.~Gottscho, T.~B. Jablin, G.~Kurian, J.~Laudon, S.~Li, P.~Ma, X.~Ma, T.~Norrie, N.~Patil, S.~Prasad, C.~Young, Z.~Zhou, and D.~Patterson, ``Ten lessons from three generations shaped google’s tpuv4i : Industrial product,'' in \emph{2021 ACM/IEEE 48th Annual International Symposium on Computer Architecture (ISCA)}, 2021, pp. 1--14.

\bibitem{jouppi2023tpu}
N.~P. Jouppi, G.~Kurian, S.~Li, P.~Ma, R.~Nagarajan, L.~Nai, N.~Patil, S.~Subramanian, A.~Swing, B.~Towles, C.~Young, X.~Zhou, Z.~Zhou, and D.~Patterson, ``Tpu v4: An optically reconfigurable supercomputer for machine learning with hardware support for embeddings,'' 2023.

\bibitem{TPUv2}
\BIBentryALTinterwordspacing
N.~P. Jouppi, D.~H. Yoon, G.~Kurian, S.~Li, N.~Patil, J.~Laudon, C.~Young, and D.~A. Patterson, ``A domain-specific supercomputer for training deep neural networks,'' \emph{Commun. {ACM}}, vol.~63, no.~7, pp. 67--78, 2020. [Online]. Available: \url{https://doi.org/10.1145/3360307}
\BIBentrySTDinterwordspacing

\bibitem{krishna2020data}
T.~Krishna, H.~Kwon, A.~Parashar, M.~Pellauer, and A.~Samajdar, ``Data orchestration in deep learning accelerators,'' 2020.

\bibitem{maestro}
H.~Kwon, P.~Chatarasi, V.~Sarkar, T.~Krishna, M.~Pellauer, and A.~Parashar, ``Maestro: A data-centric approach to understand reuse, performance, and hardware cost of dnn mappings,'' \emph{IEEE Micro}, vol.~40, no.~3, pp. 20--29, 2020.

\bibitem{kwon2018maeri}
H.~Kwon, A.~Samajdar, and T.~Krishna, ``{MAERI: Enabling Flexible Dataflow Mapping over DNN Accelerators via Reconfigurable Interconnects},'' in \emph{Proceedings of the 23rd International Conference on Architectural Support for Programming Languages and Operating Systems (ASPLOS)}, 2018.

\bibitem{CGRA}
\BIBentryALTinterwordspacing
L.~Liu, J.~Zhu, Z.~Li, Y.~Lu, Y.~Deng, J.~Han, S.~Yin, and S.~Wei, ``A survey of coarse-grained reconfigurable architecture and design: Taxonomy, challenges, and applications,'' \emph{ACM Comput. Surv.}, vol.~52, no.~6, oct 2019. [Online]. Available: \url{https://doi.org/10.1145/3357375}
\BIBentrySTDinterwordspacing

\bibitem{over_gen}
\BIBentryALTinterwordspacing
S.~Liu, J.~Weng, D.~Kupsh, A.~Sohrabizadeh, Z.~Wang, L.~Guo, J.~Liu, M.~Zhulin, R.~Mani, L.~Zhang, J.~Cong, and T.~Nowatzki, ``Overgen: Improving fpga usability through domain-specific overlay generation,'' in \emph{Proceedings of the 55th Annual IEEE/ACM International Symposium on Microarchitecture}, ser. MICRO '22.\hskip 1em plus 0.5em minus 0.4em\relax IEEE Press, 2023, p. 35–56. [Online]. Available: \url{https://doi.org/10.1109/MICRO56248.2022.00018}
\BIBentrySTDinterwordspacing

\bibitem{flexagon}
\BIBentryALTinterwordspacing
F.~Muñoz-Martínez, R.~Garg, J.~L. Abellán, M.~Pellauer, M.~E. Acacio, and T.~Krishna, ``Flexagon: A multi-dataflow sparse-sparse matrix multiplication accelerator for efficient dnn processing,'' 2023. [Online]. Available: \url{https://arxiv.org/abs/2301.10852}
\BIBentrySTDinterwordspacing

\bibitem{parashar2019timeloop}
A.~Parashar, P.~Raina, Y.~S. Shao, Y.-H. Chen, V.~A. Ying, A.~Mukkara, R.~Venkatesan, B.~Khailany, S.~W. Keckler, and J.~Emer, ``{Timeloop: A Systematic Approach to DNN Accelerator Evaluation},'' in \emph{Proceedings of the International Symposium on Performance Analysis of Systems and Software (ISPASS)}, 2019.

\bibitem{sigma_eq}
E.~Qin, A.~Samajdar, H.~Kwon, V.~Nadella, S.~Srinivasan, D.~Das, B.~Kaul, and T.~Krishna, ``Sigma: A sparse and irregular gemm accelerator with flexible interconnects for dnn training,'' in \emph{2020 IEEE International Symposium on High Performance Computer Architecture (HPCA)}, 2020, pp. 58--70.

\bibitem{constrainted_dataflow_accelerator}
J.~Seo, J.~Tong, T.~Krishna, and H.~Kwon, ``Exploring constrained dataflow accelerators for real-time multi-task multi-model ml workloads,'' in \emph{2025 IEEE International Symposium on Performance Analysis of Systems and Software (ISPASS)}, 2025, pp. 1--11.

\bibitem{squareloop}
\BIBentryALTinterwordspacing
J.~Strzeszynski, J.~Tong, K.~Lee, N.~Xiong, A.~Parashar, J.~S. Emer, T.~Krishna, and M.~Yan, ``Squareloop: Explore optimal authentication block strategy for ml,'' in \emph{Proceedings of the 14th International Workshop on Hardware and Architectural Support for Security and Privacy}, ser. HASP '25.\hskip 1em plus 0.5em minus 0.4em\relax New York, NY, USA: Association for Computing Machinery, 2025, p. 37–45. [Online]. Available: \url{https://doi.org/10.1145/3768725.3768732}
\BIBentrySTDinterwordspacing

\bibitem{tong2025MORPH}
J.~Tong, J.~Dang, S.~Langowski, T.~Huang, A.~Ali, J.~Kun, S.~Devadas, and T.~Krishna, ``Morph: Enabling ai asics for zero knowledge proof,'' in \emph{Proceedings of the 63nd Annual ACM/IEEE Design Automation Conference}, ser. DAC '26.\hskip 1em plus 0.5em minus 0.4em\relax IEEE Press, 2026.

\bibitem{tong2025CROSS}
J.~Tong, T.~Huang, J.~Dang, L.~de~Castro, A.~Itagi, A.~Golder, A.~Ali, J.~Jiang, J.~Kun, Arvind, G.~E. Suh, and T.~Krishna, ``Leveraging asic ai chips for homomorphic encryption,'' in \emph{2026 IEEE International Symposium on High Performance Computer Architecture (HPCA)}, ser. HPCA'26, Australia, 2026.

\bibitem{tong2024FEATHER}
J.~Tong, A.~Itagi, P.~Chatarasi, and T.~Krishna, ``Feather: A reconfigurable accelerator with data reordering support for low-cost on-chip dataflow switching,'' in \emph{Proceedings of the 51th Annual International Symposium on Computer Architecture}, ser. ISCA '24.\hskip 1em plus 0.5em minus 0.4em\relax Argentina: Association for Computing Machinery, 2024.

\bibitem{DSAGen}
J.~Weng, S.~Liu, V.~Dadu, Z.~Wang, P.~Shah, and T.~Nowatzki, ``Dsagen: Synthesizing programmable spatial accelerators,'' in \emph{2020 ACM/IEEE 47th Annual International Symposium on Computer Architecture (ISCA)}, 2020, pp. 268--281.

\end{thebibliography}
%%%%%%%%%%%%%%%%%%%%%%%%%%%%%%%%%%%%

\begin{table*}[t]
\centering
\small
\renewcommand{\arraystretch}{1.15}
\caption{FEATHER+ Mapper Search Knobs for a Symbolic GEMM ${\color{myred}{O}}_{m,n}^{P=M,Q=N} = {\color{myblue}{I}}_{m,k}^{M=M,J=K} \cdot {\color{mygreen}{W}}_{k,n}^{K=K,N=N}$.}
  \vspace{-3mm}

\begin{tabular}{c p{2.8cm} p{4.55cm} p{6.8cm}}
  \hline
  \textbf{Category} & \textbf{Knob} & \textbf{Symbol / Choice Set} & \textbf{Meaning} \\
  \hline

  \multirow{2}{*}{\makecell{Dataflow choice\\ $\mathrm{WO-S}, \mathrm{IO-S}$}}
  &$\mathrm{WO-S}$
  & $(M_s,K_s,N_s) = (M,K,N)$
  & Search is performed on the original GEMM. \\
  &$\mathrm{IO-S}$
  & $(M_s,K_s,N_s) = (N,K,M)$
  & Swaps the outer dimensions before search. \\

  \hline
  \multirow{3}{*}{\makecell{Tiling choice\\ (Decide \\ Partition factors)}}
  & Output-row Rank
  & $M_t \in \{AH,2AH,4AH,\dots, M_s\}$
  & Number of output rows processed per tile. \\
  & Reduction Rank
  & $K_t \in \{AH,2AH,4AH,\dots, K_s\}$
  & Reduction depth processed per tile. \\
  & Output-column Rank
  & $N_t \in \{1,2,4,\dots, N_s\}$
  & Number of output columns processed per tile. \\

  \hline
  \multirow{3}{*}{Layout choice}
  & weights order% + $N_{L0}$
  & $o_{Wvn}\!\in\!\{0,1,2,3,4,5\}$%, $N_{L0}\!\leq\!AW$
  & \tabref{tab:vn-orderid-all} permutation used for \wvn{} layout. \\
  & inputs order% + $M_{L0}$
  & $o_{Ivn} \in \{0,1,2,3,4,5\}$%, $M_{L0}\leq AW$
  & \tabref{tab:vn-orderid-all} permutation used for \ivn{} layout. \\
  & outputs order% + $P_{L0}$
  & $o_{Ovn} \in \{0,1,2,3,4,5\}$%, $P_{L0}\leq AW$
  & \tabref{tab:vn-orderid-all} permutation used for \ovn{} layout. \\

  \hline
  \multirow{5}{*}{\makecell{Combined \\ VN group \\ CG($K_g$, $N_t$)}}
  & Duplication factor
  & $d \in \left\{1,2,\dots,\left\lfloor AW / n_{\mathrm{col}} \right\rfloor\right\}$
  & Replicates each column type to many columns. \\
  \hdashline
  &  \makecell[l]{\wvn{}($k_g$, $n_t$) \\ inter-column \\ $n_t$ stride}
  & \makecell[l]{$\in \{\mathrm{block},\, \mathrm{strided}\}$}
  & \makecell[l]{Block: $s_r\!=\!1,\, s_c\!=\!AH$.
    Strided: $s_r\!=\!n_{\mathrm{col}},\, s_c\!=\!1$.\\
    Changes concurrent WVN bank access pattern.\\
    Only differs when $n_{\mathrm{col}}>1$.} \\
  \hdashline
  & \makecell[l]{\ivn{}($m_t$, $k_g$) \\ inter-column \\ $m_t$ stride}
  & \makecell[l]{$\in \{\mathrm{interleaved},\, \mathrm{consecutive}\}$}
  & \makecell[l]{Interleaved: $s_m\!=\!n_{\mathrm{rep}}$ (strided rows).\\
    Consecutive: $s_m\!=\!1$ (contiguous blocks).\\
    Changes concurrent IVN bank access pattern.\\
    Only differs when $n_{\mathrm{rep}}>1$.} \\
  \hline
  \end{tabular}
  \vspace{-5mm}
\label{tab:minisa_symbolic_knobs}
\end{table*}

\appendix
\section{Artifact Appendix}
%%%%%%%%%%%%%%%%%%%%%%%%%%%%%%%%%%%%%%%%%%%%%%%%%%%%%%%%%%%%%%%%%%%%%
\subsection{Abstract}

This artifact contains the MINISA toolchain for the FEATHER+ reconfigurable accelerator.
It includes: (1)~the FEATHER+ mapper that searches the optimal {mapping, layout} choice for processing GEMM workloads under $AH\times AW$-FEATHAER+,
(2)~a cycle-accurate analytical performance model with a 5-engine asynchronous execution simulator,
(3)~a GUI to illustrate how FEATHER+ works with cycle-by-cycle animation
(4)~instruction compression analysis comparing MINISA against explicit micro-configuration, and
(5)~latency comparison against GPU (NVIDIA RTX~5090) and TPU (Google TPU~v6e-8) baselines.
The artifact reproduces all evaluation figures: instruction reduction ratios, speedup over micro-instruction,
latency breakdown with compute utilization, and FEATHER+ vs.\ GPU/TPU latency comparison.

\subsection{Artifact check-list (meta-information)}

{\small
\begin{itemize}
  \item {\bf Algorithm: } MINISA ISA compilation for matrix multiplication on reconfigurable accelerators; Virtual Neuron (VN) abstraction with brute-force tiling search over choies in \tabref{tab:minisa_symbolic_knobs}.
  \item {\bf Program: } Python scripts: \texttt{evaluate.py} (mapping-layout cosearch), \texttt{analyze.py} (analysis + GPU/TPU comparison), \texttt{figure\_drawer/*.py} (plot generation)
  \item {\bf Compilation pipeline:}  Workload $\rightarrow$ VN groups $\rightarrow$ combined VN groups $\rightarrow$ duplication $\rightarrow$ MINISA trace $\rightarrow$ latency.
  \item {\bf Dataset: } 50 GEMM workloads from three domains: Fully Homomorphic Encryption (FHE), Zero-Knowledge Proofs (ZKP), and ChatGPT-class open-source LLM inference; pre-collected GPU/TPU baseline latency measurements
  \item {\bf Run-time environment: } Linux (Ubuntu 22.04); Python 3.10+; Conda environment with numpy, pandas, matplotlib, pyyaml
  \item {\bf Hardware: } Any x86-64 or ARM64 machine with $\geq$16\,GB RAM. No GPU/TPU required (baseline data is pre-collected). We also provide GPU and TPU scripts.
  \item {\bf Run-time state: } Deterministic analytical model (no random).
  \item {\bf Execution: } Supporting multiple-threading (\texttt{--jobs <n>})
  \item {\bf Metrics: } Compute utilization (\%), cycle breakdown (load-in, load-weight, compute, out-to-stream, store-out, instruction fetch), instruction bytes (MINISA vs.\ micro-instruction), instruction reduction ratio, latency ($\mu$s)
  \item {\bf Output: } 18 PDF figures, 6 CSV data files (benchmark summary, instruction comparison, GPU/TPU comparison, utilization/reduction/memory summaries)
  \item {\bf Experiments: } 50 workloads $\times$ 9 array configurations = 450 evaluation points (Stage~1); GPU/TPU comparison across all workloads (Stage~2); 21 publication-quality figures (Stage~3)
  \item {\bf How much disk space required (approximately)?: } $\sim$500\,MB
  \item {\bf How much time is needed to prepare workflow (approximately)?: } 5--10 minutes (conda environment setup)
  \item {\bf How much time is needed to complete experiments (approximately)?: } 3--10 hours for full evaluation (Stage~1); $<$2 minutes for analysis and plots (Stages~2--3)
  \item {\bf Publicly?: } \url{https://github.com/maeri-project/FEATHER}
  \item {\bf Archived?: } https://doi.org/10.5281/zenodo.18921947
\end{itemize}
}

%%%%%%%%%%%%%%%%%%%%%%%%%%%%%%%%%%%%%%%%%%%%%%%%%%%%%%%%%%%%%%%%%%%%%
\subsection{Description}
The MINISA framework runs on any device supporting Python. Please refer to the README.md in the open-sourced Github repository or archived artifacts DOI to install the MINISA framework, and run (1) mapping-layout cosearching, (2) layout-constrainted mapping search, (3) comparison against GPU / TPU. We provide 50 GEMM workloads as benchmark suites, as discussed in \secref{sec:workload}. We also provide pre-collected results on GPUs and TPUs.

%%%%%%%%%%%%%%%%%%%%%%%%%%%%%%%%%%%%%%%%%%%%%%%%%%%%%%%%%%%%%%%%%%%%%
\subsection{Evaluation and expected results}

(1)~the FEATHER+ mapper that searches the optimal {mapping, layout} choice for processing GEMM/convolution workloads under $AH\times AW$-FEATHAER+,
(2)~a cycle-accurate analytical performance model with a 5-engine asynchronous execution simulator,
(3)~a GUI to illustrate how FEATHER+ works with cycle-by-cycle animation
(4)~instruction compression analysis comparing MINISA against explicit micro-configuration, and
(5)~latency comparison against GPU (NVIDIA RTX~5090) and TPU (Google TPU~v6e-8) baselines.

(1) (mapping, layout) co-search for all 50 workloads under 9 configs: \texttt{python -m minisa evaluate}

(2) FEATHER+ GUI: \texttt{python -m minisa gui}

(3) vs. TPU/GPU, \texttt{python -m minisa analyze}

(4) Instruction overhead comparison (MINISA vs. micro-instruction baseline): \texttt{python -m minisa compare}

(5) (mapping, layout) co-search for GEMM/conv. under $AH\!\times\!AW$-FEATHAER+: \texttt{python -m minisa search}

(6) layout-constrainted mapping search (GEMM/conv.) under $AH\times AW$-FEATHAER+: \texttt{python -m minisa search --layout-constrained}

(7) plotting figures, \texttt{python -m minisa plot}

%%%%%%%%%%%%%%%%%%%%%%%%%%%%%%%%%%%%%%%%%%%%%%%%%%%%%%%%%%%%%%%%%%%%%
\subsection{Experiment customization}

\begin{itemize}
  \item \textbf{Array configurations:} Change \texttt{--ah} and \texttt{--aw} to evaluate different FEATHER+ sizes. Use \texttt{--aw same} for square arrays (AW=AH).
  \item \textbf{Workloads:} Supply a different CSV with different workloads. \texttt{--csv} with columns \texttt{category}, \texttt{name}, \texttt{M}, \texttt{K}, \texttt{N}.
\end{itemize}

\subsection{Mapper Design Space}
We summarize the mapper design space in \tabref{tab:minisa_symbolic_knobs} and apply heuristics to prune unpromising candidates early, accelerating the search. For 50 workloads with $AH{=}AW{=}16$, joint mapping and layout search completes in 17 minutes on an M5 Pro powered MacBook Pro using 16 parallel jobs.

% We adopt brute force search in the mapper over the provided design knobs to figure out the optimal mapping. To restrict overall design space growth.

% \subsection{\texttt{ExecuteStreaming} Example}

% As an example, consider an $AH \!\times\! 4$ PE array with the paired
% \texttt{ExecuteMapping} parameters chosen as
% $(r_0,G_r,G_c)=(0,2,1)$, and \texttt{ExecuteStreaming} set to
% $(m_0,s_m,T)=(0,3,3)$. PE columns $0$ and $1$
% belong to reduction group $j=0$, while PE columns $2$ and $3$ belong to
% reduction group $j=1$. Within each two-column group, the two columns take
% different IVN-stream phases. Therefore, over three injection cycles, the IVNs
% entering the top of the four PE columns are:
% \[
% \begin{array}{c|c|c|c|c}
% 2 \downarrow & \text{\ivn}(6,0) & \text{\ivn}(7,0) & \text{\ivn}(6,1) & \text{\ivn}(7,1) \\ 
% 1 \downarrow & \text{\ivn}(3,0) & \text{\ivn}(4,0) & \text{\ivn}(3,1) & \text{\ivn}(4,1) \\
% 0 \downarrow & \text{\ivn}(0,0) & \text{\ivn}(1,0) & \text{\ivn}(0,1) & \text{\ivn}(1,1) \\
% \hline
% \text{cycle } t & \text{PE col }0 & \text{PE col }1 & \text{PE col }2 & \text{PE col }3 \\
% \end{array}
% \]
% In the above array, Each column corresponds to one PE column in hardware, while each row denotes the \ivn{} injected at the top of that PE column in one cycle. After injection, each \ivn{} propagates downward through the corresponding PE column over consecutive cycles, so the same streamed operand is temporally reused by all PEs in that column.

\end{document}